\documentclass[acmsmall,screen,nonacm]{acmart}

\setcopyright{none}

\usepackage{enumitem}
\usepackage{caption}
\usepackage{mymacros}
\usepackage{lang}
\usepackage{hyperref}
\usepackage{todonotes}
\usepackage{float}
\usepackage{subcaption}
\usepackage{multirow}
\usepackage{url}
\usepackage{wrapfig}

\newcommand{\SOLIDITYKEYWORDS}

\lstnewenvironment{solidity}
    {\lstset{%
      name=main
      language=Java,         
      xrightmargin=2pt,
      xleftmargin=10pt,
      columns=fullflexible,
      keepspaces=true,
      style=numbers,
      classoffset=0,
      frame=bt,
      deletekeywords={byte, char, int, float, double},
      keywordstyle=\color{blue}\bfseries,      
      classoffset=1,
      morekeywords={modifier, mapping, contract, require, uint, emit,
                    assert, if, else, contract, function, returns, var, public}, 
      keywordstyle=\color{brown}\bfseries,  
      classoffset=0,
      commentstyle=\color{blue},             
      morecomment=[l][\color{mauve}]{//},
      morecomment=[l][\color{red}]{//!},
      morecomment=[s][\color{mauve}]{/*}{*/},
      morecomment=[s][\color{red}]{/**}{*/},
      stringstyle=\color{dkgreen},             
      escapeinside={\%*}{*)}                   
}}{}

\lstnewenvironment{solidity-small}
    {\lstset{%
      name=main
      language=Java,         
      columns=fullflexible,
      keepspaces=true,
      style=tinynonumbers,
      classoffset=0,
      frame=bt,
      deletekeywords={byte, char, int, float, double},
      keywordstyle=\color{blue}\bfseries,      
      classoffset=1,
      morekeywords={modifier, mapping, contract, require, uint, emit,
                    assert, if, else, contract, function, returns, var, public}, 
      keywordstyle=\color{brown}\bfseries,  
      classoffset=0,
      commentstyle=\color{blue},             
      morecomment=[l][\color{mauve}]{//},
      morecomment=[l][\color{red}]{//!},
      morecomment=[s][\color{mauve}]{/*}{*/},
      morecomment=[s][\color{red}]{/**}{*/},
      stringstyle=\color{dkgreen},             
      escapeinside={\%*}{*)}                   
}}{}

\AtBeginDocument{%
  }

\begin{document}

\title{The Incredible Shrinking Context... in a Decompiler Near You}

\author{Sifis Lagouvardos}
\orcid{0000-0002-6233-1548}
\affiliation{%
  \institution{University of Athens}
  \city{Athens}
  \country{Greece}
}
\affiliation{%
  \institution{Dedaub}
  \city{Athens}
  \country{Greece}
}
\email{sifis.lag@di.uoa.gr}

\author{Yannis Bollanos}
\orcid{0009-0006-6905-9264}
\affiliation{%
  \institution{Dedaub}
  \city{Athens}
  \country{Greece}
}
\email{ybollanos@dedaub.com}

\author{Neville Grech}
\orcid{0000-0002-6790-2872}
\affiliation{%
  \institution{Dedaub}
  \city{Msida}
  \country{Malta}
}
\email{me@nevillegrech.com}

\author{Yannis Smaragdakis}
\orcid{0000-0002-0499-0182}
\affiliation{%
  \institution{University of Athens}
  \city{Athens}
  \country{Greece}
}
\affiliation{%
  \institution{Dedaub}
  \city{Athens}
  \country{Greece}
}
\email{smaragd@di.uoa.gr}



\bibliographystyle{ACM-Reference-Format}

\tolerance=8000

\widowpenalty=6000
\clubpenalty=6000

\newcommand{\ourtool}{\textsc{Shrnkr}}
\newcommand{\heimdall}{\textsc{Heimdall-rs}}


\newcommand{\AppendixA}{Appendix~\ref{sec:cloningFull}}
\newcommand{\AppendixB}{Appendix~\ref{sec:incompleteGlobalFull}}
\newcommand{\AppendixC}{Appendix~\ref{app:study}}
\newcommand{\Appendices}{\newpage
\appendix

\section{Control Flow Normalization via Cloning}\label{sec:cloningFull}

A second technique that enables precision in \ourtool{} is the aggressive cloning of blocks that are \emph{locally}
determined to be used in inconsistent ways. (\emph{Local} inspection refers to inspection that does not require
the full power of the decompiler's static analysis, i.e., the shrinking context of Section~\ref{sec:shrinking}.
Effectively, the inspection examines the block's contents and occurrences of the block address constant in the contract
code.)

\subsection{Cloning Need: Illustration}
As mentioned in Section~\ref{sec:background}, the Solidity compiler will often try to reuse the same low-level blocks for different high-level purposes.
Such an optimization is essential in the setting of smart contracts as the Ethereum blockchain enforces a size limit of 24576 bytes~\cite{eip-170}.

The code snippet of the earlier Figure \ref{fig:calls:chained} exhibits such behavior with block \sv{0x72},
which is pushed to the stack twice as a continuation, in the course of performing two of the three checked subtraction
operations (per line 5 of our earlier Solidity example of Figure~\ref{fig:solidity-example}).

For this example, the state-of-the-art Elipmoc decompiler will correctly identify that this block is used to perform two different function calls, attempting to summarize them as
shown in the three-address-code snippet in Figure \ref{fig:chained:elipmoc}.

\begin{figure}[]
  \centering
  \vspace{-0.2cm}
  \begin{subfigure}[b]{0.95\textwidth}
    \includegraphics[width=\linewidth]{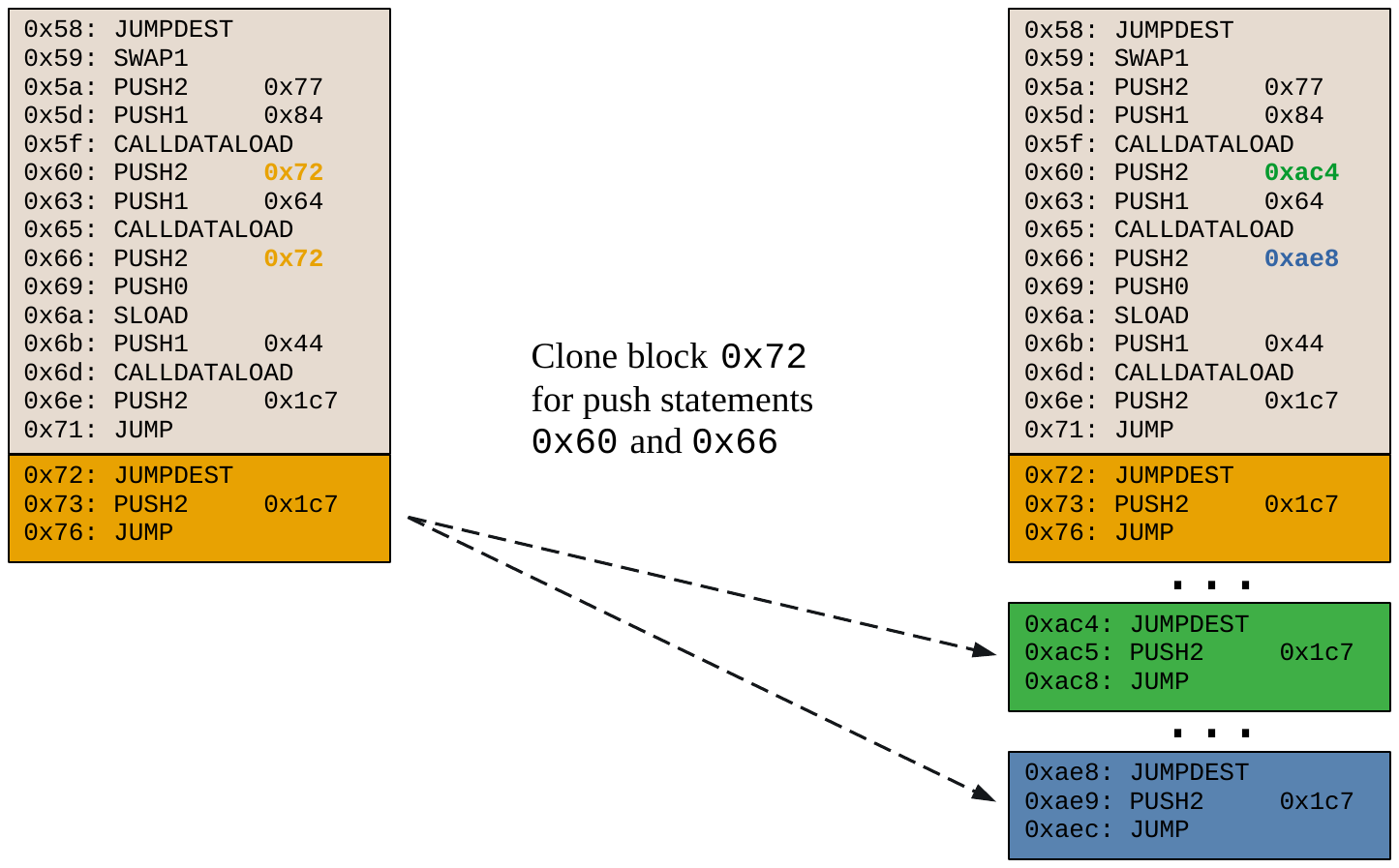}
    \vspace{-2.2cm}\\
    \caption{Cloning transformation: block-level view}
      \label{fig:chained:illustration}
  \end{subfigure}
  \begin{subfigure}[b]{0.5\textwidth}
      \centering
      \begin{EVMnonumbercodesmallwbox}
Begin block 0x58
prev=[0x178B0x50], succ=[0x72]
=================================
0x5a: v5a(0x77) = CONST
0x5d: v5d(0x84) = CONST
0x5f: v5f = CALLDATALOAD v5d(0x84)
0x60: v60(0x72) = CONST
0x63: v63(0x64) = CONST
0x65: v65 = CALLDATALOAD v63(0x64)
0x66: v66(0x72) = CONST
0x69: v69 = PUSH0
0x6a: v6a = SLOAD v69
0x6b: v6b(0x44) = CONST
0x6d: v6d = CALLDATALOAD v6b(0x44)
0x6e: v6e(0x1c7) = CONST
0x71: v71_0 =CALLPRIVATE v6e(0x1c7), v6d, v6a, v66(0x72)

Begin block 0x72
prev=[0x72, 0x58], succ=[0x72, 0x77]
=================================
0x72_0x0: v72_0 = PHI v71_0, v76_0
0x72_0x1: v72_1 = PHI v47(0x20), v5f, v65, v150V44
0x72_0x2: v72_2 = PHI v60(0x72), v44(0xab), v5a(0x77), v166V50
0x73: v73(0x1c7) = CONST
0x76: v76_0 = CALLPRIVATE v73(0x1c7), v72_0, v72_1, v72_2 // last argument is the continuation
        \end{EVMnonumbercodesmallwbox}
      \caption{Elipmoc Output}
      \label{fig:chained:elipmoc}
  \end{subfigure}%
  ~
  \begin{subfigure}[b]{0.5\textwidth}
      \centering
      \begin{EVMnonumbercodesmallwbox}
Begin block 0x58
prev=[0x50], succ=[0xae8]
=================================
0x5a: v5a(0x77) = CONST
0x5d: v5d(0x84) = CONST
0x5f: v5f = CALLDATALOAD v5d(0x84)
0x60: v60(0xac4) = CONST
0x63: v63(0x64) = CONST
0x65: v65 = CALLDATALOAD v63(0x64)
0x66: v66(0xae8) = CONST
0x69: v69(0x0) = CONST
0x6a: v6a = SLOAD v69(0x0)
0x6b: v6b(0x44) = CONST
0x6d: v6d = CALLDATALOAD v6b(0x44)
0x6e: v6e(0x1c7) = CONST
0x71: v71_0 = CALLPRIVATE v6e(0x1c7), v6d, v6a, v66(0xae8)

Begin block 0xae8
prev=[0x58], succ=[0xac4]
=================================
0xae9: vae9(0x1c7) = CONST
0xaec: vaec_0 = CALLPRIVATE vae9(0x1c7), v71_0, v65, v60(0xac4)

Begin block 0xac4
prev=[0xae8], succ=[0x77]
=================================
0xac5: vac5(0x1c7) = CONST
0xac8: vac8_0 = CALLPRIVATE vac5(0x1c7), vaec_0, v5f, v5a(0x77)
        \end{EVMnonumbercodesmallwbox}

      \caption{\ourtool{} Output}
      \label{fig:chained:shrinker}
  \end{subfigure}
  \caption{Cloning illustration and decompilation output of optimized chained call pattern using Elipmoc (i.e., without cloning support) and \ourtool{}}
\end{figure}

The results of this merging of different high-level calls are detrimental to the precision of downstream analyses.
Following the function call at block \sv{0x72}, the control-flow is transferred either to itself or to block \sv{0x77}.
(Note this in the \sv{succ} annotation in the block's header.)
The Elipmoc output in Figure \ref{fig:chained:elipmoc} simply cannot be expressed using standard linear IR control-flow constructs
over the basic blocks shown: the decompilation output needs to explicitly refer to passed continuations to explain the successor
blocks.

Such artifacts of imprecise decompilation output have been identified in past literature~\cite{elipmoc} as ``Unstructured Control Flow''.

In addition, the identities of the arguments of the two function calls are not recoverable.
Through the introduction of phi functions, the first argument (\sv{v72\_0}) is the merging of two variables
and the second one (\sv{v72\_1}) the merging of 4 variables.
As a result, an analysis that intends to precisely model the effects of the high-level calls will need to consider 8
combinations of arguments, 6 of which are invalid.

Both of the above problems are addressed in the output of \ourtool{} in Figure~\ref{fig:chained:shrinker}.
This is achieved via the low-level cloning transformation shown in \ref{fig:chained:illustration}:
For each of the two low-level statements pushing block \sv{0x72} to the stack, we generate a new cloned copy of
the block and replace their pushed value with the identifier of their corresponding fresh block.

To improve the precision and completeness of our decompilation output we employ heuristics to identify low-level blocks that are likely
reused to implement more than one high- or low-level construct and clone all their different uses.

\subsection{Identifying Block Cloning Candidates}

We will now describe the two block categories we identified as candidates for cloning.
It should be noted that both of these block categories describe small basic blocks with few to no
high-level operations in them.

\myparagraph{Reused Continuations}
The first class of blocks we identify as candidates for cloning are blocks that are used as continuations in
more than one case (i.e., by more than one push statement). These continuation blocks are often used
to perform chained calls at different points in a contract's execution (as in the example in Figure \ref{fig:calls:chained})

\myparagraph{Stack Balancing Blocks}
The second class of blocks we attempt to clone are \emph{Stack-Balancing Blocks}.
These are blocks that contain only low-level stack-altering instructions (i.e., \sv{POP}, \sv{SWAPx}, \sv{DUPx})
and end with a \sv{JUMP}.
An example \emph{Stack Balancing Block} can be seen in the following EVM code snippet:
\begin{EVMnonumbercode}
  0x1c3: JUMPDEST
  0x1c4: SWAP1
  0x1c5: POP
  0x1c6: SWAP2
  0x1c7: SWAP1
  0x1c8: POP
  0x1c9: JUMP
\end{EVMnonumbercode}

\emph{Stack Balancing Blocks} are \emph{locally-unresolved} blocks that are typically highly-reused in a single contract.
These blocks are often used to implement the return blocks of different private functions or even used in different ways
inside the same private function.

\subsection{Block Cloning Details}

After identifying the block-cloning candidates, we find the ones that are used in more than one place in the low-level bytecode
(i.e., are pushed to the stack by more than one bytecode statement).
We encode our block-cloning instances as tuples of [pushStmt, blockToClone] and generate a new low-level block for each tuple.
To reduce implementation complexity we only allow the cloning of low-level blocks that end with \sv{JUMP} statements,
having no fallthrough block that would need to be cloned as well.

\section{Incomplete Global Pre-Analysis}\label{sec:incompleteGlobalFull}

The general framework of shrinking context sensitivity admits several
refinements.
The last new technique from \ourtool{} that we present plays
multiple roles, e.g., in eliminating spurious call edges as well as in
introducing more edges than mere calls into the context.  Both cases
require a global pre-analysis, i.e., perform a best-effort,
\emph{incomplete} version of the full decompiler static analysis and
use it to configure the subsequent complete analysis.

\subsection{Overview}

The main context-sensitive global control-flow analysis is preceded by an \emph{incomplete} global pre-analysis round.
The pre-analysis helps in the following aspects of the decompiler.

\myparagraph{Spurious \predname{PrivateCallAndContinuation}}
Recall the \predname{PrivateCallAndContinuation} predicate we used when defining our context
sensitivity model. The default meaning of this predicate would be as-defined in prior work:
these \emph{likely} function calls are identified in the global analysis model of
the Gigahorse~\cite{gigahorse} decompiler, which is also used by the Elipmoc~\cite{elipmoc} decompiler.
Specifically:
\begin{bullets}
  \item A \emph{likely} function call is a block that jumps to another block directly, leaving
    the stack with one or more \emph{likely} continuations on it.
  \item A value pushed to stack is considered a \emph{likely} continuation if a \sv{JUMPDEST}
    exists at that index.
\end{bullets}

Thus if a block leaves the stack with a value that happens to be the same as a \sv{JUMPDEST} but it does not end up
being used as a jump target, our \textbf{Merge} constructor will add that \emph{spurious} caller to the calling context.
\ourtool{} refines this definition with a much deeper pre-analysis step.

\myparagraph{Spurious \predname{PublicCall}}
The definition of our \emph{shrinking context sensitivity} model also relies on the \predname{PublicCall} predicate,
locating the entries of the contract's public functions.
As the results of the \predname{PublicCall} are utilized by its global control-flow graph analysis,
Elipmoc uses local heuristic patterns to identify public function entries.

These patterns identify comparisons with small constant values (like \sv{0x12e49406} and \sv{0x87d7a5f4} in
our example in Figure \ref{fig:pubFunn}) that are \emph{likely} to be function selector values.
However one cannot confirm that these values are compared against the function selector bytes
loaded from the call-data using only a local analysis.
Such imprecise \predname{PublicCall} inferences can have a negative impact on our context-sensitive
control-flow graph.

\myparagraph{Imprecision-Introducing Edges}
The heuristics used in shrinking context sensitivity aim to approximate private function calls and returns.
However there can be non-call edges that would make our global analysis more precise if they were to be included in the calling
context. The decompiler can include a generic, agnostic (to control-flow idioms, such as ``calls'' and ``returns'')
way to identify such edges, based on blocks that introduce global
imprecision.

\subsection{Implementation}

The implementation of the incomplete global analysis is rather straightforward; we halt the execution of the global
analysis after a set number of facts have been produced.

\ourtool{} is implemented as a set of Datalog inferences, using the Souffle~\cite{Jordan16, DBLP:conf/cc/ScholzJSW16,souffleInterpreted} engine.
We restrict the global pre-analysis to a bounded amount of work by using souffle's \sv{.limitsize} directive
\footnote{\url{https://souffle-lang.github.io/directives\#limit-size-directive}} on the key \predname{BlockOutput} relation.
This directive will halt the bottom-up evaluation of predicate \predname{BlockOutput} once its size limit is reached. (We use a
coarse limit of 1,000,000 facts, a number derived empirically, but not particularly tuned.)
In addition, reaching the limit will also halt the evaluation of all other mutually-recursive relations in the same stratum, essentially stopping
the incomplete global pre-analysis, as we intend.

The inference logic of the incomplete global pre-analysis uses core relations in the
decompiler. These are summarized in Figure~\ref{fig:schema2} but are essentially a model that was inherited
from past static-analysis-based decompilers, tracing back to Gigahorse~\cite{gigahorse}.

\begin{figure*}[tb!p]
\begin{small}
  \begin{tabular}{l}
  $V$: set of virtual ``variables'', i.e., instructions that push values or stack positions at block entry  \\
  $B$: set of basic blocks \\
  $PC$: set of private contexts, \args{PC} $\cong$ $B^n$ \\
  $C$: set of contexts, \args{C} $\cong$ \args{B $\times$ PC}, as labeled record \record{\args{B}}{\args{PC}}\\
  \end{tabular}

\begin{tabular}{l l l}
\cline{1-2}
\pred{BlockInput}{ctx: C, block: B, i: $\mathbb{N}$, var: V} & \multirow{2}{*}{$\begin{cases} \shortstack[l]{At entry/exit to/from \args{block}, under context \args{ctx},\\
  the \args{i}-th position of the stack contains \args{v}.} \end{cases}$}\\
\pred{BlockOutput}{ctx: C, block: B, i: $\mathbb{N}$, var: V} &  \\
\pred{BlockJumpTarget}{ctx: C, from: B, var: V, to: B} &  Block \args{from}, under context \args{ctx}, pushes \args{var} on the \\
  &  stack, whose value is a jump target, \args{to}.\\
\pred{GlobalBlockEdge}{fromCtx: C, from: B, toCtx: C, to: B} & Block \args{from} under context \args{fromCtx}\\
  &  may jump to block \args{to} under context \args{toCtx}.\\
\cline{1-2}
\noalign{\vskip 1mm}
\end{tabular}
\end{small}
\caption[]{Core relations in decompiler, used in global pre-analysis.}
\label{fig:schema2}
\end{figure*}

\subsection{Removing Spurious PrivateCallAndContinuation facts}

The incomplete global pre-analysis helps answer the question ``is the jump at the end
of this block really a call?'' based on global behavior, rather than local heuristics,
as in past work.  That is, the global pre-analysis simply simulates the static analysis
up to bounded work and summarizes the observed behaviors. We filter out the spurious \predname{PrivateCallAndContinuation} inferences
by ensuring
the variable holding the value of the continuation is used as a jump target for \emph{some} block and context,
as shown in the rule below (with \predname{Pre\_PrivateCallAndContinuation} being the original definition):

\begin{datalogcode}
  PrivateCallAndContinuation(block, continuation):-
    Pre_PrivateCallAndContinuation(block, continuation),
    BlockPushesContinuation(block, contVar, continuation),
    BlockJumpTarget(_, _, contVar, continuation).
\end{datalogcode}


If our global analysis work limit is too small, causing some valid execution behaviors to not be modeled,
some valid \predname{PrivateCallAndContinuation} facts could be considered spurious
and be filtered out. This could negatively affect the precision and scalability of the
main global analysis: the context abstraction could be missing elements, collapsing together behaviors
that would be best kept distinct. The rest of our pipeline will remain unaffected.
It is up to our experimental evaluation to validate that the incompleteness
of the global analysis does not practically limit the effectiveness of the main analysis.

\subsection{Removing Spurious PublicCall facts}

We use the output of the incomplete global pre-analysis to filter out any spurious
\predname{PublicCall} inferences similarly to how we filtered spurious \predname{PrivateCallAndContinuation}
facts in the previous subsection.

In the case of \predname{PublicCall} facts we introduce a relation that tracks the variables
that hold the function selector bytes, and ensure that one of these variables is used in a comparison
with a small (<= 4 bytes) value.

\subsection{Recovering Important Edges}

The most interesting application of the global pre-analysis is
for identifying imprecision-introducing edges in a \emph{generic, agnostic} way.
This is captured in the definition of predicate \predname{ImportantEdge}, in Figure
\ref{fig:importantEdges}.  Effectively, the logic offers a generic way
to assess that a block is crucial for precision: it is reached with
the stack contents having multiple distinct values (at some position),
and none of its predecessors have this property upon their exit. That
is, the block is reached by multiple predecessors, each establishing
their own run-time conditions. Thus, the edge reaching the
block is important for precision, and therefore worth remembering in the
static context.

\begin{figure}
\begin{datalogcode}
  ImpreciseBlockInput(ctx, block, index):-
    BlockInput(ctx, block, index, var1),
    BlockInput(ctx, block, index, var2),
    var1 != var2.

  ImpreciseBlockOutput(ctx, block, index):-
    BlockOutput(ctx, block, index, var1),
    BlockOutput(ctx, block, index, var2),
    var1 != var2.

  ImpreciseBlockInputFromPrevious(ctx, block, index):-
    ImpreciseBlockInput(ctx, block, index),
    GlobalBlockEdge(prevCtx, prevBlock, ctx, block),
    ImpreciseBlockOutput(prevCtx, prevBlock, index).

  ImprecisionIntroducedAtEdge(fromCtx, fromBlock, toCtx, to, index):-
    GlobalBlockEdge(fromCtx, fromBlock, toCtx, to),
    ImpreciseBlockInput(toCtx, to, index),
    !ImpreciseBlockInputFromPrevious(toCtx, to, index),
    !ImpreciseBlockOutput(fromCtx, fromBlock, index).

  ImportantEdge(from, to):-
    ImprecisionIntroducedAtEdge(_, from, _, to, _).
\end{datalogcode}
\caption{Logic identifying imprecision introducing block edges}
\label{fig:importantEdges}
\end{figure}

The introduction of \predname{ImportantEdge} requires additions to our \funcname{Merge} constructor, which can be
found in Figure~\ref{fig:context-sensitivity-rules-v2}.

\begin{figure}
  \begin{tabular}{l}
  $B$: set of basic blocks \\
  $PC$: set of private contexts, \args{PC} $\cong$ $B^n$ \\
  $C$: set of contexts, \args{C} $\cong$ \args{B $\times$ PC}, as labeled record \record{\args{B}}{\args{PC}}\\
  \\
  Initially, \args{ctx} = \record{\textsc{Null}}{[]}\\
\funcname{Merge}(\record{\args{u}}{\args{p}}, \args{cur}, \args{next}) =
$\begin{cases}
    \record{\args{next}}{p}, \ \ \text{if \pred{PublicCall}{cur, next}}\\
    \\[-1em]

    \record{\args{u}}{[\args{cur}, \predname{First}_{n-1}(\args{p})]},\\
    \begin{tabular}{l} \text{if \pred{PrivateCallAndContinuation}{cur, *}} \\
      \text{or (\pred{PrivateReturn}{cur}} \\
    \ \ \ \  \text{and ($\nexists c$ $\in \args{p}$: \pred{PrivateCallAndContinuation}{c, next})}\\
      \text{or (\pred{ImportantEdge}{cur, next})}
    \end{tabular} \\

    \record{\args{u}}{\funcname{CutTo}(\args{p}, c)},\\
    \begin{tabular}{l}
      \text{if \pred{PrivateReturn}{cur}}\\
      \ \ \ \  \text{and ($\exists c$ $\in \args{p}$: \pred{PrivateCallAndContinuation}{c, next})}
    \end{tabular} \\
    \\[-1em]
    \record{\args{u}}{\args{p}},  \text{otherwise} \\
  \end{cases}
  $

\end{tabular}
\caption[]{Context constructor for shrinking context
  sensitivity following an incomplete global pre-analysis.}
  \label{fig:context-sensitivity-rules-v2}
\end{figure}

In the case of the pre-analysis incompleteness having a negative impact, we will miss some possible \predname{ImportantEdge} inferences,
without affecting our main global analysis or the rest of our pipeline in any way. Again, this effect will be evaluated
experimentally.

\section{Human Study Details}
\label{app:study}

In this Appendix we'll present the layout of the form completed by the human study participants.

\subsection{Invitation}

You've been invited to participate in a study conducted by researchers at the University of Athens and Dedaub on the quality of EVM bytecode decompilers.
The study evaluates three state-of-the-art industrial and academic decompilers.
You will be assigned 3 tasks, each designed to take 10-20 minutes.

For each task you will be given the source code of a smart contract along with the decompilation outputs of the 3 tools, and a method.
You will be tasked with evaluating the decompilation outputs for the given method and report which of the tools are able to accurately recover the method's logic.

To the extend possible, you should attempt to ignore the differences in decompilation style such as the inference of private functions
vs the inlining of all code, naming conventions, etc.

\subsection{Demographic Questions}

\begin{enumerate}
  \item How many years of experience do you have in EVM smart contract security and/or binary reverse engineering?
  \begin{itemize}
    \item 0--2
    \item 2--4
    \item >=5
  \end{itemize}
  \item Which of the options best describes your role?
  \begin{itemize}
    \item Security Researcher/Auditor
    \item Reverse Engineer
    \item Student
    \item Smart Contract Developer
    \item Other
  \end{itemize}
\end{enumerate}

\subsection{Decompilation Task}

\begin{enumerate}
  \item Please share the details of your assigned task (link to gist and requested method).
  \item Decompiler A was able to to accurately express the logic of the given method.
  \begin{itemize}
    \item Strongly Disagree
    \item Disagree
    \item Neutral
    \item Agree
    \item Strongly Agree
  \end{itemize}
  \item Decompiler B was able to to accurately express the logic of the given method.
  \begin{itemize}
    \item Strongly Disagree
    \item Disagree
    \item Neutral
    \item Agree
    \item Strongly Agree
  \end{itemize}
  \item Decompiler C was able to to accurately express the logic of the given method.
  \begin{itemize}
    \item Strongly Disagree
    \item Disagree
    \item Neutral
    \item Agree
    \item Strongly Agree
  \end{itemize}
\end{enumerate}}

\begin{abstract}
  Decompilation of binary code has arisen as a highly-important
  application in the space of Ethereum VM (EVM) smart contracts. Major
  new decompilers appear nearly every year and attain popularity, for
  a multitude of reverse-engineering or tool-building purposes.
  Technically, the problem is fundamental: it consists of recovering
  high-level control flow from a highly-optimized continuation-passing-style (CPS)
  representation. Architecturally, decompilers can be built using
  either static analysis or symbolic execution techniques.

  We present \ourtool{}, a static-analysis-based decompiler succeeding
  the state-of-the-art Elipmoc decompiler. \ourtool{} manages to achieve drastic
  improvements relative to the state of the art, in all significant
  dimensions: scalability, completeness, precision. Chief among the
  techniques employed is a new variant of static analysis context:
  \emph{shrinking context sensitivity}. Shrinking context sensitivity
  performs deep cuts in the static analysis context, eagerly
  ``forgetting'' control-flow history, in order to leave room for
  further precise reasoning.

  We compare \ourtool{} to state-of-the-art decompilers,
  both static-analysis- and symbolic-execution-based. In a standard benchmark set,
  \ourtool{} scales to over 99.5\% of contracts (compared to $\sim$95\% for Elipmoc),
  covers (i.e., reaches and manages to decompile) 67\% more code than \heimdall{},
  and reduces key imprecision metrics by over 65\%, compared again to Elipmoc.

\end{abstract}

\maketitle

\section{Introduction}

\emph{Decompilation} or \emph{lifting} from low-level binary code to a structured, high-level representation is a problem with a
substantial history and practical significance in a variety of settings~\cite{quteprints36820,StaticDisassembly,javaDecompSurvey,harrand2019strengths}.
In the context of programmable blockchains, decompilation has found a new application domain, with difficult
technological considerations but intense demand. \emph{Smart contracts} (the colloquial name for
programs on a programmable blockchain) are deployed publicly and executed by-consensus of the entire network.
Decompiling smart contracts is in high demand for several applications: building automated analyses
over a uniform representation (regardless of the existence or not of source code for the smart contract);
reverse-engineering security attacks (where source code is unavailable); understanding competitive trading
strategies by trading bots (where source code is unavailable); and much more.

The dominant binary platform for smart contracts is the Ethereum VM
(EVM). It is the execution layer for most programmable blockchains,
such as Ethereum, BSC, Arbitrum, Polygon, Optimism, Fantom, Base,
Avalanche, and more. Accordingly, the problem of decompiling EVM
bytecode has received significant attention~\cite{gigahorse,elipmoc,vandal,heimdall,panoramix,EVMDecompStudy} and new
entrants constantly vie for adoption---e.g., with the recent \heimdall{} repo~\cite{heimdall}
rapidly reaching 1,000 stars and 100 forks.

From a technical standpoint, the problem of EVM decompilation is especially challenging. The EVM bytecode language is extremely low-level
with respect to control flow, replacing all execution-control constructs with jump instructions to an address popped from the
execution stack. That is, all control-flow statements (e.g., conditionals, loops, function calls, function returns) are translated
into jumps to an address that is a run-time value of the low-level program.
The challenge of EVM decompilation, thus, is to derive a higher-level representation, including
functions, calls, returns, and structured control flow, from EVM bytecode. As a program analysis
challenge, it has been the domain for applying several techniques. The primary distinction is
between \emph{symbolic execution} approaches~\cite{panoramix,heimdall} and \emph{static analysis}~\cite{gigahorse,elipmoc} approaches.
Symbolic-execution-based decompilers are easier to develop, naturally produce partial results (i.e., can always
produce \emph{something}, rather than failing on a smart contract in its entirety), yet are often vastly incomplete, failing
to even discover a significant portion of the code. In contrast, static-analysis-based decompilers typically require much
heavier development effort, cover (nearly) all code, but can fail to scale or can produce imprecise output.

In this work, we present a static-analysis-based decompiler that significantly advances the state of the art, on
all quality dimensions (precision, completeness, scalability). Compared to Elipmoc~\cite{elipmoc}, its predecessor and the leading
static-analysis-based decompiler, our tool, \ourtool{}, achieves much greater scalability (up to 99.7\% on Elipmoc's evaluation
dataset, compared to Elipmoc's 95.3\%), while substantially
improving precision and completeness---virtually nullifying imprecision or incompleteness for most metrics.
Compared to the modern, most-adopted symbolic-execution-based decompiler, \heimdall{}, \ourtool{} exhibits a
large advantage, decompiling up to 67\% more binary statements. In essence, for complex contracts, \ourtool{} succeeds in
decompiling much of the interesting logic, while \heimdall{} simply fails to find values to even cover the
deepest statements via \emph{one} path, let alone via all the different paths that can lead to such statements.

The technical essence of \ourtool{} lies in several improvements over past static-analysis-based approaches:
more precise and scalable static modeling, control-flow normalization via cloning, and pre-analysis-guided
elimination of spurious calls. One key novelty is responsible for the lion's share of the benefit: the fundamental static model
is improved, by use of a new kind of static \emph{context}
kept inside the decompiler. That is, the decompiler maintains as its current control-flow history (i.e., how the
execution got to the currently-analyzed statement) a list of basic blocks that is updated under a different
algorithm. The new logic, dubbed \emph{shrinking context sensitivity}, aggressively \emph{shrinks} the context
when a likely matching call-return or chained-call pattern is observed.

In overview, the key contributions of this work consist of:
\begin{itemize}
\item a new algorithmic specification of context sensitivity, shrinking context sensitivity, suitable for the domain
  of EVM smart contract decompilation;
\item an array of other techniques (block cloning, incomplete global pre-analysis to prepare the main analysis)
  that contribute to precision, completeness, and scalability;
\item an experimental evaluation demonstrating substantial improvement over past decompilers in all interesting
  axes.
\end{itemize}


\section{Background}
\label{sec:background}

We next introduce the setting of this work: smart contracts, decompilation, context sensitivity.

\subsection{EVM Smart Contracts}

Smart contracts are small programs (typically up to around 1,000 lines of high-level code, limited to 24KB in binary form in Ethereum) stored on a persistent blockchain as part of its state.
They are typically written in a high-level programming language, with Solidity being by far the most widely used.
Solidity, which is the setting of our work, dominates over all other languages in terms of adoption.
At the high-level, a smart contract defines a set of external/public functions, which are its public entry points through which
Externally-Owned Accounts (\emph{EOA}s) or other smart contracts can interact with the contract, and a set of persistent \emph{storage} variables which
are part of the contract's state on the blockchain. Code reuse is facilitated through the use of \emph{internal} (a.k.a. \emph{private}) functions,
inheritance, and library contracts. Solidity is a statically-typed language supporting operations on a number of value types
(signed and unsigned integers, bytes, boolean), dynamic-length arrays, and associative mappings,
as well as complex types combining the above.

The execution setting of smart contracts exhibits several intricacies, many of which are relevant
to our discussion of bytecode analysis and decompilation.
Performing transactions on the EVM requires a \emph{gas} fee, paid in the chain's native token.
This cost is (intended to be) analogous to the effort the blockchain's nodes need to perform, and I/O-heavy tasks (e.g.,
random access to blockchain state) are much more costly.
As a result, a smart-contract compiler will typically optimize for two things: decreasing the executable bytecode's size,
and reducing the runtime gas cost of its transactions.
As a virtual machine, the EVM is powerful but simple and very low-level. It is a stack-based machine that
supports arithmetic and logic operations over 256-bit (32-byte) words, has an implicitly persistent heap area
(called \emph{storage}), and a transient heap-like area (called \emph{memory}). Types, objects, functions, closures,
arrays, records, and any other high-level concepts are all translated away into word-level operations at the EVM level.
This means that operations for most data types will require additional low-level code performing bit shifting or masking.

In the EVM, basic blocks are explicitly delineated, via \sv{JUMPDEST} and \sv{JUMP}/\sv{JUMPI} (collectively: \emph{jump}) instructions.
The flow between blocks, however, is far from clear.
The EVM's jump statements are inherently dynamic, reading the value of the target block from the stack.\footnote{In this paper,
the term \emph{block} refers to a \emph{basic block}, as in standard compilers literature, i.e., a maximal sequence of low-level
instructions always executed from start to finish. This has no connection to the ``block'' in ``blockchain''. Since our work
does not involve distributed systems considerations, we never need to refer to the latter.}
Although most jump targets can be \emph{resolved locally} (i.e., by looking at each basic block in isolation),
the existence of \emph{locally unresolved} dynamic jumps makes the computation of the control-flow-graph (CFG) an involved task.
Each transaction involving a contract begins at statement \sv{0x0} and goes on until a statement that halts execution is reached.
The EVM offers no primitives for defining and calling functions, requiring the use of low-level code patterns to support
public and private functions.

In addition, compiler version and settings greatly affect the produced bytecode.
The release of Solidity v0.8.0~\cite{solidity08} introduced checked arithmetic and employed v2 of the ABI encoder,
greatly increasing the number of internal functions.
Since Solidity v0.8.13~\cite{solidity0813}, a new compilation pipeline became stable,
with plans to make it the default in a future release~\cite{solidityYulBlog}.
This new \emph{Yul/viaIR} pipeline involves Yul: a standardized, exportable
intermediate language/representation (IR), which is also integrated into the Solidity language as inline assembly~\cite{ChaliasosAssembly}.
The Yul/viaIR pipeline enables deeper optimizations and more auditable code generation
than the currently default ``legacy'' pipeline.

\subsection{Context Sensitivity}

Context sensitivity~\cite{Shivers1991,dvanhorn:milanova-etal-tosem05,10.1145/1925844.1926390,dvanhorn:emami-etal-pldi94,dvanhorn:Sharir:Interprocedural} has a long history in static program analysis, with work in over 3 decades. Every static analysis
that computes the flow of abstract values through the program is trying to approximate the solution to an undecidable 
problem. As a result, it faces challenges with respect to both scalability and precision. 
Addressing such challenges requires careful design decisions and context-sensitivity has offered a good way to balance these
needs.

Context-sensitivity associates program variables (and sometimes heap objects) with context information and distinguishes the
values not just on the basis of variables but on the basis of variable+context combinations.
Analysis inferences for multiple executions that result in the same context will be merged,
but stay differentiated from inferences associated with different contexts.
Call-site-sensitivity, i.e. using one or more previous call-sites as context information, has seen success in analyzing functional languages~\cite{Shivers1991}
and low-level imperative languages\cite{dvanhorn:emami-etal-pldi94}.
For object-oriented languages, the use of the receiver object(s) as context information has been the state-of-the-art context sensitivity abstraction
since its introduction. Section~\ref{sec:related} offers more detailed pointers and comparison with past work.

\subsection{EVM Decompilation}
We define the problem of EVM bytecode decompilation (a.k.a. \emph{binary lifting})
as the derivation of high-level control-flow
constructs and program structure from EVM bytecode. One can view the problem as the
attempt to reconstruct a high-level program\footnote{Notably, none of
the EVM decompilers produce code that can be re-compiled.
This does not diminish the value of EVM
decompilation: the output of state-of-the-art decompilers is typically
excellent both for human consumption and for writing automated program
processing tools (e.g., static analyzers~\cite{symvalic}, symbolic-execution
tools~\cite{ConfusumContractum, NotYourType}, or program
verification engines~\cite{Grossman2017, 9784829, TamingCallbacks}).}
from a low-level, stack-based intermediate
representation (IR), where all control flow is represented in a
\emph{continuation-passing style (CPS)} form. For instance, a function
call is done by pushing a continuation on the stack (the address of
the basic block to return to), then pushing the function's entry block
address, and jumping. All control-flow patterns, such as in-function
branching, tail calls, calls in-sequence, passing a return value of a
call as an argument to another, etc., are represented as complex
sequences of pushing continuations and eventually jumping to the
first.

In the setting of EVM decompilation, the dynamic nature of the \sv{JUMP} operations creates the need for whole-program
reasoning, in order to compute a program's control-flow graph. The Gigahorse/Elipmoc framework~\cite{gigahorse,elipmoc} has addressed this need
by introducing a global context-sensitive control-flow graph/points-to analysis as the backbone of its decompiler.

This context-sensitive global control-flow graph is then used by the Elipmoc framework~\cite{elipmoc}, which we extend, to identify
potential call-sites, which are in turn used to compute function boundaries. Lastly, after the function boundaries are
computed, their number of arguments and return arguments are inferred.

The Gigahorse lifter employed a N-call-site (or jump-site) context-sensitivity algorithm, while Elipmoc introduced a composite approach
that included the identity of the public entry point and the 8 last call-sites that are likely private function calls or returns.
The evaluation of the Elipmoc publication highlighted the key importance the context-sensitivity algorithm plays in the decompiler's
scalability and precision.

\section{Motivation: Solidity to EVM by example}

We next showcase various elements of binary-level EVM smart contracts
as produced by the Solidity compiler. These motivate and provide important
context for our later discussion.

\subsection{Compiler Translation}
To glimpse the low-level complexity of compiled smart contracts, we consider a simple example.
The program of Figure~\ref{fig:solidity-example} contains two external functions that accept various parameters and perform
fund-transfer operations. As we discuss later, the expression \sv{amt - defaultFee - feeA - feeB}
on line 5 actually performs three function calls to a private function used to check subtraction
for underflow.

\begin{figure}[h]
\begin{solidity}
interface IERC20 { function transfer(address to, uint256 value) external returns (bool); }
contract DecomptTest {
  uint256 defaultFee;
  function transWFee(address tok, address to, uint256 amt, uint256 feeA, uint256 feeB) external
  { IERC20(tok).transfer(to, amt - defaultFee - feeA - feeB); /* 3 private function calls */ }
  function simpleTransfer(address tok, address to, uint256 amt) external
  { IERC20(tok).transfer(to, amt); }
}
\end{solidity}
\caption{Simple smart contract, used as running example.}
\label{fig:solidity-example}
\end{figure}

From the perspective of decompilation, the Solidity compiler is two different compilers, because of
the aforementioned Yul/viaIR pipeline. The compiler effectively has two entirely separate
code generation back-ends, which produce vastly different binary code patterns. Different
optimization levels also greatly affect the binary program. Observable high-level metrics,
such as the bytecode size, or internal metrics, such as the number of private/internal
functions (which are not apparent in the final binary but are a key concept in the intermediate compiler
representations) vary greatly, as shown for an example contract in the table below.

\begin{small}
\begin{center}
\begin{tabular}{|r|c|c|c|c|c}
  \hline
  Compiler Configuration & Bytecode Size & Number of Internal functions \\
  \hline
  legacy, no optimizer & 1,000 & 20  \\
  viaIR, no optimizer & 1,195 & 43 \\
 legacy, optimizer level 200 & 667 & 8 \\
 viaIR, optimizer level 200 & 542 & 6 \\
  \hline
\end{tabular}
\end{center}
\end{small}

\subsection{Public Function Patterns}
As public functions are not inherent in the EVM, high-level languages adhere to the contract Application Binary
Interface (ABI)~\cite{abi}, which specifies how input and output data are to be encoded when interacting
with a smart contract.
Per the ABI, the first 4 bytes provided in a smart contract invocation are the \emph{function selector},
used to identify the public function being called.

\begin{figure}[tbh]
\begin{EVMnonumbercodesmall}
0x1a: CALLDATALOAD
0x1b: PUSH1     0xe0
0x1d: SHR                   // function selector === calldataload(0) >> 28
0x1e: DUP1
0x1f: PUSH4     0x12e49406
0x24: EQ
0x25: PUSH2     0x38        // push function address of transWFee() public function
0x28: JUMPI
0x29: DUP1
0x2a: PUSH4     0x87d7a5f4
0x2f: EQ
0x30: PUSH2     0x54        // push function address of simpleTransfer() public function
0x33: JUMPI    ...
\end{EVMnonumbercodesmall}
\caption{Function selector logic for our example in Figure \ref{fig:solidity-example}.}
\label{fig:pubFunn}
\end{figure}

Figure~\ref{fig:pubFunn} shows the bytecode implementing the function selector logic for our example.
These compiler-produced patterns implementing the function-selector logic have been used by past tools~\cite{gigahorse,elipmoc,Albert2018}
to identify public function entries. However even detecting such simple patterns can have challenges.
Looking at the previous code segment it is trivial for a local analysis to deduce that the \sv{EQ} statement \sv{0x24}
operates on the function selector, since the input variable is loaded in the same block. \sv{EQ} statement \sv{0x2f} also
checks the function selector, however an inter-block analysis is required to be able to deduce this.
The state-of-the-art Elipmoc binary lifter requires computing the public function entries before performing its
global control-flow graph analysis~\cite[Figure 4]{elipmoc}. In order to achieve this, it uses an approximation based on a local-only analysis that does not
verify that the selector is actually used.

In Section~\ref{sec:incompleteGlobal} we propose a 2-phase global analysis that, among others, tackles this problem.

\subsection{Private Function Patterns}
\label{sec:private-functions}

The absence of native support for internal/private functions for the EVM has led to the emergence of low-level patterns to
support code reuse.

\begin{figure*}[tbh]
  \begin{subfigure}[b]{0.48\textwidth}
      \centering
      \begin{EVMnonumbercodesmallwbox}
0x12a: PUSH2  0x132  // push continuation address
0x12d: DUP3          // position argument in stack
0x12e: PUSH2  0x109  // push function address
0x131: JUMP
0x132: JUMPDEST      // continuation address
...
0x109: JUMPDEST      // cleanup_t_uint160 function address
0x10b: PUSH20  0xffffffffffffff...ffffffffffffffffff
0x121: AND           // masks arg's upper 12 bytes off
0x124: SWAP2         // shuffles stack
0x127: JUMP          // jumps to continuation
\end{EVMnonumbercodesmallwbox}
      \caption{Simple Private Function Call}
      \label{fig:calls:simple}
  \end{subfigure}%
  ~
  \begin{subfigure}[b]{0.48\textwidth}
      \centering
      \begin{EVMnonumbercodesmallwbox}
0x58: JUMPDEST
0x59: SWAP1
0x5a: PUSH2  0x77   // pushes final continuation
0x5d: PUSH1  0x84
0x5f: CALLDATALOAD
0x60: PUSH2  0x72   // pushes cont. for 3rd call
0x63: PUSH1  0x64
0x65: CALLDATALOAD
0x66: PUSH2  0x72   // pushes cont. for 2nd call
0x69: PUSH0
0x6a: SLOAD
0x6b: PUSH1  0x44
0x6d: CALLDATALOAD
0x6e: PUSH2  0x1c7  // push safeSub func address
0x71: JUMP
0x72: JUMPDEST      // cont. address for 2nd, 3rd call
0x73: PUSH2  0x1c7  // push safeSub func address
0x76: JUMP
0x77: JUMPDEST      ...
\end{EVMnonumbercodesmallwbox}

      \caption{Optimized Chained Private Function Calls}
      \label{fig:calls:chained}
  \end{subfigure}
  \caption{Private Function Call Patterns}
\end{figure*}

The basic pattern for private function calls, presented in Figure~\ref{fig:calls:simple}, has been identified in past literature~\cite{gigahorse,elipmoc,NeuralFEBI}.
A basic block makes a call to internal function \sv{cleanup\_t\_uint160} at offset \sv{0x109},
having first pushed the bytecode offset at which it wants to return after the called function's execution
completes. The return block of a function is a \emph{locally unresolved} block that jumps back
to the continuation block pushed by its caller.

In case of optimized code, a block performing an internal function call can also push the continuations of future calls.
This pattern is typically produced in cases of calls that can be chained together (such as complex arithmetic expressions).
As an example, consider the optimized compilation of expression `\sv{amt - defaultFee - feeA - feeB}' from our example
program. The subtraction operations are actually function calls, to a checked-subtraction function, so the expression
should be thought of as `\sv{safeSub(safeSub(safeSub(amt,defaultFee),feeA),feeB)}'.

The resulting bytecode can be seen in Figure~\ref{fig:calls:chained}.

In this optimized case, block \sv{0x58} will set the stack so that the 3 checked sub-operations are chained.
Earlier work on the Elipmoc decompiler includes a function reconstruction algorithm~\cite[Figure 6]{elipmoc} that recursively infers these
chained function calls. It assumes that each low-level block will map to one high-level function call.
However, in optimized code, the same low-level block can be used to perform more than one high-level function call.
This can be seen in the above example, which pushes the address of block \sv{0x72} on the stack twice,
implementing in this way the last two checked-subtraction operations.
Effectively, if we think of our high-level code as `\sv{safeSub$_1$(safeSub$_2$(safeSub$_3$(amt,defaultFee),feeA),feeB)}'
then \sv{safeSub$_2$} and \sv{safeSub$_3$} are a \emph{single} low-level instruction, the same for both calls.

Such complex patterns evade past function reconstruction algorithms. In Section \ref{sec:cloning} we propose a cloning-based technique that identifies
such low-level blocks with different high-level uses, and clones them, recovering precision of decompilation output.

\subsection{Our Context}

\ourtool{} is the third iteration of the Gigahorse/Elipmoc~\cite{gigahorse,elipmoc} lifter framework, building on the foundation of the state-of-the-art Elipmoc~\cite{elipmoc} tool.
Thus, \ourtool{} is available as an open source tool on the public repository of the Gigahorse framework.\footnote{\url{https://github.com/nevillegrech/gigahorse-toolchain/tree/sub24}}

Elipmoc's combination of scalability, precision, and completeness, paired with its expressive IR, have
established it as a dominant lifter for EVM bytecode.
Research tools for diverse program analysis applications have been implemented on top of Elipmoc. The applications include
static-analysis~\cite{ethainter,ethereum-memory-oopsla,DefiTainter,SmartDagger,symvalic,Grech2020cacm,TODLER,10.1145/3597503.3639153,PrettySmart,PROXYEX,StorageModeling},
symbolic execution~\cite{NotYourType,ConfusumContractum}, and deep learning~\cite{DeepInfer}.

\ourtool's novel techniques mainly improve the decompiler's context-sensitive global
control-flow graph with the introduction of the \emph{shrinking context sensitivity} variant described in
Section \ref{sec:shrinking} and its tuning via incompleteness in Section \ref{sec:incompleteGlobal}.
Section \ref{sec:cloning} describes the introduction of a block cloning transformation step, which is performed before the
global analyses and helps \ourtool{} produce normalized decompilation output.

Apart from the above, \ourtool{} inherits Elipmoc's architecture, design decisions, and implementation
with the exception of shallow fixes. The main components inherited from Elipmoc are its componentized
local analyses, function reconstruction algorithms, and IR generation pipeline.

\section{Shrinking Context-Sensitivity}\label{sec:shrinking}

We next present the main algorithmic techniques that help our tool, \ourtool{}, drastically improve
over the state of the art in EVM decompilation. Chief among them is \emph{shrinking context sensitivity}, a new
analysis context abstraction.

Past work~\cite{elipmoc, gigahorse} has established a context-sensitive global control-flow-graph analysis as the backbone of a decompiler.
That is, the decompiler abstractly simulates all possible executions of the decompiled program, but in a finite space: instead
of keeping a full, unbounded execution stack, the decompiler collapses the stack into a finite \emph{context} structure. That is, both
the dynamic execution stack and the static context can be thought of as sequences of basic blocks, with the static
context being a bounded sequence.

The essence of the context-sensitivity algorithm is to decide \emph{which} elements of the execution stack to keep at every point of
modification, i.e., at every jump instruction. Different dynamic executions that have the same context (because their differing
elements have been dropped by the context-sensitivity algorithm) will be treated the same, with the analysis computing all
possible values for a variable, instead of just a single value.

As demonstrated by the Elipmoc work~\cite{elipmoc} the choice of context sensitivity algorithm greatly affects a decompiler's scalability and output quality.
In contrast to the N-call-site sensitivity employed by Gigahorse~\cite{gigahorse}, Elipmoc proposed a
\emph{transactional context-sensitivity} variant consisting of two parts: a sticky public function component,
and a private function context including the $N$ latest likely private function calls or returns.
The publication's evaluation confirmed that both of the components of \emph{transactional context-sensitivity}
had a positive impact on scalability and precision.

Our approach, dubbed \emph{shrinking context sensitivity}, retains the two-part approach with a key distinction:
the private function context can \emph{shrink} (much more drastically than merely discarding the oldest element),
disregarding the context elements that are related
to a likely private call, after that call returns to the first continuation pushed by its caller.

\begin{figure}
  \includegraphics[trim={0 1.5cm 0 2.5cm},clip,width=1.03\linewidth]{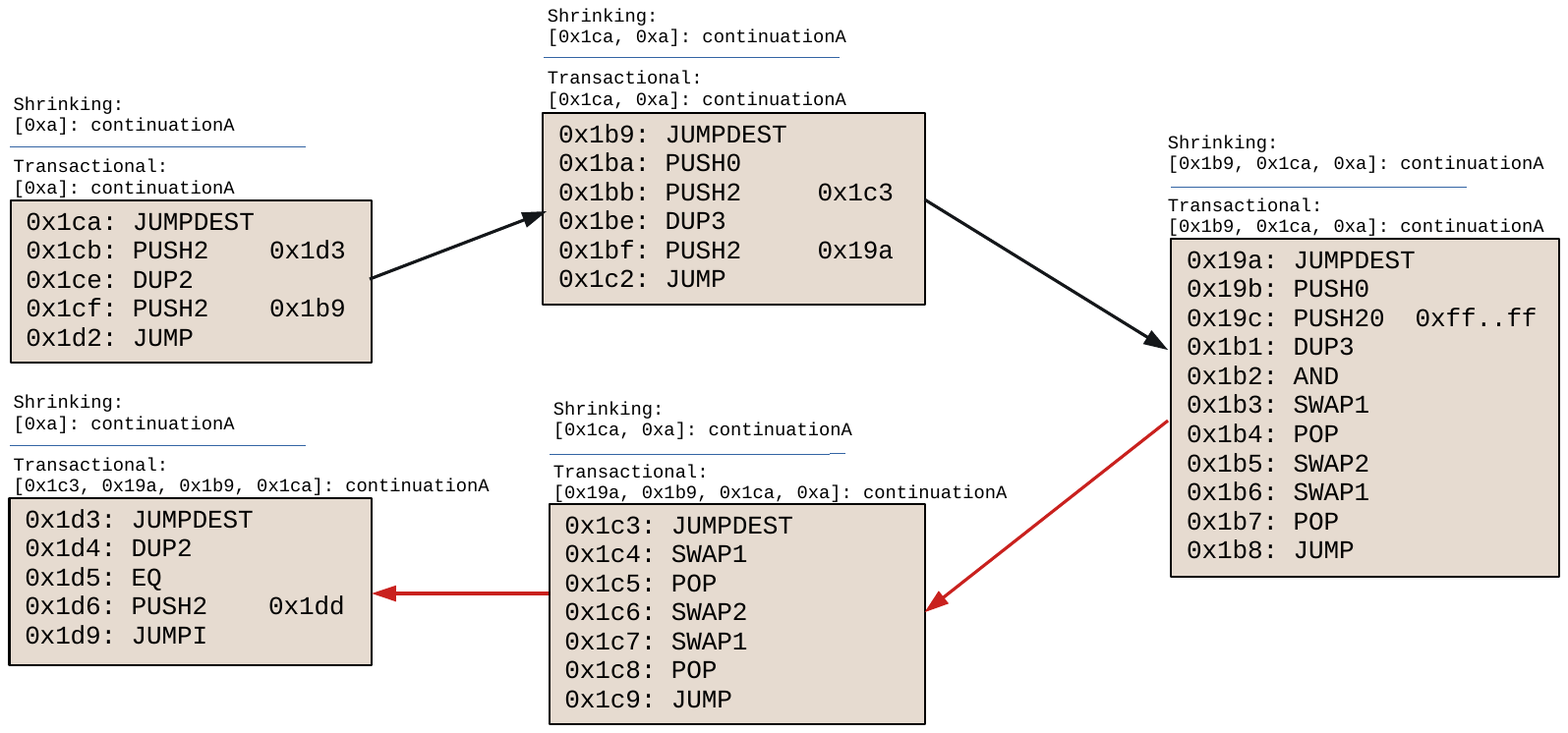}
  \caption{Example: \emph{Shrinking} context sensitivity contrasted (at each analyzed block) to
    the \emph{Transactional} context sensitivity of past work. Both context sensitivity algorithms have
    a maximum context depth of 4. The public function components of both algorithms are omitted because
    they remain unchanged in the transitions shown. Arrows to the right are calls, arrows to the left returns.
    The analysis has initial information that should be kept precisely through the analyzed sub-graph: \sv{continuationA}
    is applicable (e.g., it is kept in a certain stack location) if we reach the first analyzed block (\sv{0x1ca})
    with context \sv{0xa}. Transactional context sensitivity forgets this information by the time it analyzes the
    last block: the context is merely the blocks shown in the figure, with no trace of how the analysis got to them.
    In contrast, shrinking
    context sensitivity maintains the information: the context shown at the last block captures how we got to the first block.
    }
  \label{fig:shrinkExample}
  \end{figure}

The example of Figure~\ref{fig:shrinkExample} helps explain the intuition behind \emph{shrinking context sensitivity}.
The example illustrates the effects of \emph{shrinking context sensitivity}, contrasted with Elipmoc's
\emph{transactional context-sensitivity}, on the private function context, for a series of block transitions.
Blocks \sv{0x1ca} and \sv{0x1b9} likely perform function calls as, following their execution,
they leave the stack with continuations (\sv{0x1d3} and \sv{0x1c3}, respectively) pushed to it.
For \emph{shrinking context sensitivity}, when these continuations are reached, the most recent blocks, up to the
block that pushed the continuation on the stack, are dropped.
This leaves room to maintain other information, within the same maximum context depth: \emph{shrinking context sensitivity} reaches
block \sv{0x1d3} with the same calling context
as the initial block \sv{0x1ca}, retaining crucial information about how the latter was reached.
In contrast, \emph{transactional context-sensitivity} reaches its maximum context depth and has to remove the oldest
element from the stack, when it analyzes block \sv{0x1d3}. This will mean that if block \sv{0x1ca} was reachable under more
than one calling context, upon reaching block \sv{0x1d3}, \emph{transactional context-sensitivity} is unable to
differentiate these contexts, merging them all into one.

\begin{figure}
  \begin{tabular}{l}
  $B$: set of basic blocks \\
  $PC$: set of private contexts, \args{PC} $\cong$ $B^n$ \\
  $C$: set of contexts, \args{C} $\cong$ \args{B $\times$ PC}, as labeled record \record{\args{B}}{\args{PC}}\\
  \\
  Initially, \args{ctx} = \record{\textsc{Null}}{[]}\\
\funcname{Merge}(\record{\args{u}}{\args{p}}, \args{cur}, \args{next}) =
   $\begin{cases}
    \record{\args{next}}{p}, \ \ \text{if \pred{PublicCall}{cur, next}}\\
    \\[-1em]

    \record{\args{u}}{[\args{cur}, \predname{First}_{n-1}(\args{p})]},\\
    \begin{tabular}{l} \text{if \pred{PrivateCallAndContinuation}{cur, *}} \\
      \text{or (\pred{PrivateReturn}{cur}} \\
    \ \ \text{and ($\nexists c$ $\in \args{p}$: \pred{PrivateCallAndContinuation}{c, next}))}
    \end{tabular} \\

    \record{\args{u}}{\funcname{CutTo}(\args{p}, c)},\\
    \begin{tabular}{l}
      \text{if \pred{PrivateReturn}{cur}}\\
      \ \ \text{and ($\exists c$ $\in \args{p}$: \pred{PrivateCallAndContinuation}{c, next})}
    \end{tabular} \\
    \\[-1em]
    \record{\args{u}}{\args{p}},  \text{otherwise} \\
  \end{cases}
  $

\end{tabular}
\caption[]{Context constructor for shrinking context
  sensitivity.
  For ease of exposition, we use labeled records to distinguish the public part of the context
  (single element) from the private part (of $n$ elements), instead of
  merging both in a flat tuple of $n+1$ elements.}
  \label{fig:context-sensitivity-rules}
\end{figure}

\begin{figure*}[tb!p]
\begin{small}
\begin{tabular}{l l l}

\cline{1-2}
\pred{PublicCall}{cur: B, next: B} & Block transition is likely an entry to a public function.\\
\pred{PrivateCallAndContinuation}{caller: B, cont: B} & The \args{caller} block likely makes a private function call \\
& after having pushed block \args{cont} as a continuation.\\
\pred{PrivateReturn}{cur: B} & The current block likely returns from a private \\
& function call. \\
\func{CutTo}{p: PC, b: B}{$p'$} & Truncating private context \args{p} until encountering \args{b} \\
&  yields \args{$p'$}. \\
\cline{1-2}
\noalign{\vskip 1mm}
\end{tabular}
\end{small}
\caption[]{Auxiliary relations.}
\label{fig:schema}
\end{figure*}

Figure~\ref{fig:context-sensitivity-rules} presents the definition of
shrinking context sensitivity, in compact form. (A description in English follows shortly, and the reader may
choose to consult it before referring to the formal definition.) The
context-sensitivity definition is given in the form of the \funcname{Merge} context
constructor. Namely, the value
\funcname{Merge}(\record{\args{u}}{\args{p}}, \args{cur}, \args{next})
gives the analysis context for basic block \args{next} when the
analysis finds an edge (i.e., a possible jump) from basic block
\args{cur} to \args{next} and the current analysis context for block
\args{cur} is \record{\args{u}}{\args{p}}.

Figure~\ref{fig:schema} gives
definitions for auxiliary relations that we refer to both in the context-sensitivity
definition and in later logical specifications. It is important to note that function-inference
predicates such as \predname{PrivateCallAndContinuation}
and \predname{PrivateReturn} are only \emph{likely} true to their name. The analysis
cannot know for sure when a control-flow transition corresponds to a high-level
function call. At this stage, the analysis can only, at best, grossly \emph{over-approximate}
what \emph{might} be the possible calls and returns. (This over-approximation will contain
many more edges than what will be eventually deemed to be function calls and returns.)
However, the naming reflects the intuition: we want shrinking context sensitivity to
attempt to match function calls and returns, and hopefully achieve both precision and
scalability even with this incomplete information.

As a reminder, in predicate \pred{PrivateCallAndContinuation}{caller: B, cont: B},
the continuation does not have to be the block to return to after the call performed by block
\args{caller} (as it would be in a straightforward, unoptimized compilation of
a simple call). It can instead be the return block for the caller's caller (in case of tail calls),
or the entry block of another called function (in case of chained calls), or any other block
determined by complex optimizing compilation patterns.

\noindent In English, the definition of Figure~\ref{fig:context-sensitivity-rules} states:

Upon a block transition,
\begin{itemize}
\item if a public function entry is found, enter it as the public part of the context; otherwise
\item if a likely private function entry is found, push the caller block in the private part of the context,
  simultaneously dropping the oldest block in the context. Do the same if the transition is a likely
  function return that cannot be matched with a call in the context. Matching is done
  by comparing the continuation (that the earlier likely call has pushed) with the one the return
  is going to;
\item if the block transition is a likely function return and can be matched with a call in the
  context, then drop all top-most private context elements until reaching the matching call;
\item in all other cases the context is propagated as it is.
\end{itemize}

The intuition behind shrinking context sensitivity is deceivingly simple: static context abstracts away the
dynamic execution stack of the EVM. It then stands to reason that when the dynamic execution
returns from a function call, no record of the function entry should remain on the static context,
much like in the dynamic stack. The analogy is not perfect, however. First, as discussed, call and return
block transitions are far from certain. Second, in the dynamic execution
stack, it is not the caller block of a function that is kept during the call, but only the continuation (i.e.,
the code where the function will return).

These two differences play into each other. The static analysis defensively keeps limited information to deal with natural uncertainty.
(This uncertainty is due to not being certain about function call/return transitions but also due to the static
loss of precision relative to dynamic execution, because of truncating state to a bounded size.)
But it can drop information when it develops higher confidence: when the static analysis sees a \emph{likely} call, it cannot
be confident enough that it \emph{is} a call and will eventually return, thus it keeps the called block in the context.
When a matching return is found, however, the analysis confidence increases enough to remove not only the
(likely) function call block but also all other blocks pushed on the context between the function entry and the return:
these blocks are very likely intra-function control flow.

Truncating the context enables much greater precision later, since the context depth is finite (and would otherwise
need to ``forget'' potentially valuable prior state about previous blocks that led to the current one).
Additionally, the truncation logic offers a natural self-healing mechanism for the analysis abstraction of execution
context: even if some inference (i.e., determining that a block may be a call and should thus be kept in the context)
turns out to be noisy, it will likely be pruned when an enclosing function returns.

\section{Other Enhancements}

\ourtool{} also integrates some secondary enhancements compared to the Elipmoc decompiler.

\subsection{Control Flow Normalization via Cloning}\label{sec:cloning}
\ourtool{} performs aggressive cloning of blocks that are \emph{locally}
determined to be used in inconsistent ways. (\emph{Local} inspection refers to inspection that does not require
the full power of the decompiler's static analysis, i.e., the shrinking context of Section~\ref{sec:shrinking}.)

The motivation for cloning has already been discussed with the private function reconstruction example
of Section~\ref{sec:private-functions}. (A detailed example and explanation
can be found in the extended version of the paper, in \AppendixA{}.)
Effectively, the cloning transformation discovers blocks that are used as continuations in
more than one case (i.e., by more than one push statement). These continuation blocks are often used
to perform chained calls at different points in a contract's execution (as in the example in Figure \ref{fig:calls:chained}).
We encode our block-cloning instances as tuples of [pushStmt, blockToClone] and generate a new low-level block for each tuple.
To reduce implementation complexity we only allow the cloning of low-level blocks that end with \sv{JUMP} statements,
having no fallthrough block that would need to be cloned as well.

\subsection{Incomplete Global Pre-Analysis}\label{sec:incompleteGlobal}
\ourtool{} leverages a \emph{global pre-analysis} before the main decompilation step. The pre-analysis is a best-effort,
\emph{incomplete} version of the full context-sensitive control-flow graph analysis and we use it to configure the
subsequent complete analysis in the following ways:

\begin{itemize}
\item We remove spurious \sv{PublicCall} inferences by ensuring that the blocks identified by the detection of the
local patterns presented in Figure \ref{fig:pubFunn} actually operate on a stack location holding the function selector bytes.
\item We filter out spurious \sv{PrivateCallAndContinuation} inferences by making sure that the locally identified \textit{likely} private calls
push a continuation block that is actually used as a target in a subsequent \sv{JUMP} operation.
\item We identify block transitions that lead to imprecision in the global control-flow graph analysis.
\end{itemize}

The first two cases help our analysis stay more precise and scalable by ensuring precious space in the public or
private context components is not wasted on false inferences whose inclusion in the contexts give no additional precision
benefits. The final case also helps reduce imprecision by identifying important edges not captured by our local heuristic
rules.

\AppendixB{} provides additional details on the incomplete global pre-analysis.

\section{Evaluation}\label{sec:eval}

The evaluation of \ourtool{} intends to answer three distinct research questions:\\
\emph{RQ1: Comparison with static-analysis-based decompilers}
How does \ourtool{} compare against the closest comparable state-of-the-art static-analysis-based decompiler?\\
\emph{RQ2: Comparison with symbolic-execution-based decompilers}
How does \ourtool{} compare against the most popular symbolic-execution-based decompiler? \\
\emph{RQ3: Design Decisions}
How do the various technical components of \ourtool{} (Sections~\ref{sec:shrinking},~\ref{sec:cloning},~\ref{sec:incompleteGlobal})
affect its results?

\subsection{Experimental Setup}

We perform the evaluation of \ourtool{} using 2 experimental datasets:

\paragraph*{Standard Dataset}
The first dataset is that used in the publication and artifact for the state-of-the-art Elipmoc binary lifter.
The dataset consists of~\cite{elipmoc}: \emph{5,000 unique contracts, first deployed on the main Ethereum network between
blocks 12,300,000 (April 24, 2021) and 13,300,000 (September 26, 2021).}

\paragraph*{Yul Dataset}
To investigate how the different tools do on the recently-released Yul/IR pipeline
we introduce a new dataset consisting of 3,000 unique contracts compiled using the Yul/IR pipeline,
deployed on the Ethereum mainnet until block 18,750,000 (Dec 09, 2023).

Although the Yul/IR pipeline is still used for a small minority of deployed smart
contracts, it is likely to become more dominant in the future, especially after it becomes
default. (Although becoming default does not immediately signify adoption: developers in this space are particularly sensitive to compilation
settings and routinely override the defaults for deployment.) Furthermore, the Yul/IR
pipeline is explicitly much harder for decompilers.\footnote{Cf. recent comments of Solidity lead
developers: ``\emph{Decompilation is more complicated, yes}'' and ``\emph{For decompilers it could be a problem}'',
\url{https://youtu.be/3ljewa1__UM?t=921} .}

Experimental runs are performed on a machine with 2 Intel Xeon Gold 6136 12 core CPUs and 754G of RAM.
An execution cutoff of 200s was used for all tools. (This is over an order of magnitude higher than the
average decompilation time of a contract. That is, if the decompiler does not finish in 200s, it is unlikely to
ever finish, due to exponential explosion in the number of contexts, expressing failure to maintain precision.)
For \ourtool{} and Elipmoc we performed the experiments
using 24 concurrent jobs, taking advantage of their out-of-the-box support for the parallel analysis
of a set of contracts. We performed the \heimdall{} runs sequentially as it lacks such support.

When performing the evaluation, we noticed that \heimdall{} was often spending most of its execution time
querying an online database to resolve the signatures of public methods via their function selector values.
This results in a cosmetic-only improvement in the output, by showing high-level identifiers.
Thus, in order to avoid disadvantaging \heimdall{}, we used the \sv{--skip-resolving} flag when invoking it.
To match, we deleted all entries on the files \ourtool{} and Elipmoc use for the resolution of public function signatures.

Our configuration of \ourtool{} sets the maximum context depth of the \emph{shrinking context sensitivity} to 20.
In addition when we refer to Elipmoc's \emph{transactional context sensitivity} we use a maximum depth of 8,
as set in the Elipmoc publication. Elipmoc is largely unscalable with deeper context. Generally,
these parameters are chosen as defaults by the respective tool authors
because they are close to ``experimentally optimal'', so to speak. One
can change them to improve some metric (e.g., higher values will improve precision), at the expense of
others (incurring more timeouts).

\subsection{Comparison with Elipmoc}

Elipmoc~\cite{elipmoc} is
the state-of-the-art research decompiler for EVM smart contracts. It has also seen industrial success by being the core
of the infrastructure of the Dedaub Contract Library and Security Suite, available at \url{https://app.dedaub.com/}.
The shared core of \ourtool{} and Elipmoc allows us to perform an in-depth comparison.
In all numbers shown in this section, \textbf{lower is better}. That is, precision, completeness, and scalability are
evaluated via metrics of \emph{imprecision}, \emph{incompleteness}, and \emph{lack of scalability}, respectively.

\subsubsection{Scalability}

Perhaps the topmost quality axis for a static-analysis-based decompiler is how often its static model
scales well. (Without sacrificing precision, as confirmed later.)
We compare the scalability of the two tools in Table \ref{tab:elipmoc:scale}.

\begin{table}[htb]
  \caption{Timeouts of \ourtool{} and Elipmoc. \textbf{Left table}: Standard Dataset, \textbf{Right table}: Yul Dataset.}
  \begin{small}
  \begin{minipage}{.5\linewidth}
    \centering
    \begin{tabular}{c|c|}
      {} &  Timeouts \\
      \hline
      \ourtool{}  &       13 (\textbf{0.26\%}) \\
      Elipmoc     &      235 (\textbf{4.7\%}) \\
      \hline
      Total &      5000 \\
      \end{tabular}      
  \end{minipage}%
  \begin{minipage}{.5\linewidth}
    \centering
      \begin{tabular}{|c|c}
        {} &  Timeouts \\
        \hline
        \ourtool{}  &       94 (\textbf{3.13\%}) \\
        Elipmoc     &      379 (\textbf{12.63\%}) \\
        \hline
        Total &      3000 \\
      \end{tabular}
  \end{minipage}
  \end{small}
  \label{tab:elipmoc:scale}
\end{table}

\ourtool{} vastly outscales Elipmoc in both datasets:
For the Standard dataset, it manages to decompile nearly all contracts, with a timeout rate of
just 0.26\% versus Elipmoc's 4.7\%.
For the Yul dataset, the difference is again very significant with \ourtool{} achieving 3 times fewer
timeouts, at a timeout rate of 3.13\%, compared to Elipmoc's 12.63\%.

This gives us an initial confirmation on the difference of the two datasets and their underlying
code generation pipelines. The newer, more powerful Yul/IR pipeline provides a significantly increased
challenge to decompilers which we will also see confirmed later in this section.

Table~\ref{tab:elipmoc:scale:depth} breaks down this performance by size class. As can be seen, for the largest
contracts (15KB and above), Elipmoc very often fails. \ourtool{} drops the timeout rates by a factor of 3 or more
in all size classes.

\begin{table}[htb]
  \caption{\ourtool{} and Elipmoc's timeouts for each contract size class.\\
    \textbf{Top table}: Standard Dataset. \textbf{Bottom table}: Yul Dataset.}
\hspace{-1cm}  \begin{minipage}{0.8\linewidth}
  \centering
  \begin{small}
    \begin{tabular}{r|c|c|c|c|c}
    Bytecode Size & [0,5KB) & [5KB,10KB) & [10KB,15KB) & [15KB,20KB) & [20KB,max) \\
    \hline
    \ourtool{} & 2 (\textbf{0.08\%}) & 4 (\textbf{0.38\%}) & 2 (\textbf{0.31\%}) & 1 (\textbf{0.34\%}) & 4 (\textbf{0.91\%})  \\
    Elipmoc   & 5 (\textbf{0.2\%}) & 40 (\textbf{3.76\%})  & 111 (\textbf{17.1\%}) & 39 (\textbf{13.22\%}) & 40 (\textbf{9.11\%}) \\
    \hline
    Contracts in size class     & 2552          & 1065         & 649          & 295          & 439
\end{tabular}
  \end{small}
\end{minipage}
  
\vspace{0.7cm}
\hspace{-1cm}\begin{minipage}{0.8\linewidth}
  \centering
  \begin{small}
  \begin{tabular}{r|c|c|c|c|c}
  Bytecode Size & [0,5KB) & [5KB,10KB) & [10KB,15KB) & [15KB,20KB) & [20KB,max) \\
  \hline
  \ourtool{} & 2 (\textbf{0.17\%}) & 12 (\textbf{1.66\%}) & 14 (\textbf{3.35\%}) & 29 (\textbf{8.68\%}) & 37 (\textbf{10.22\%})  \\
  Elipmoc   & 17 (\textbf{1.46\%}) & 54 (\textbf{7.49\%})  & 62 (\textbf{14.83\%}) & 114 (\textbf{34.13\%}) & 132 (\textbf{36.46\%}) \\
  \hline
  Contracts in size class     & 1165          & 721         & 418          & 334          & 362
\end{tabular}
  \end{small}
\end{minipage}
\label{tab:elipmoc:scale:depth}
\end{table}

\subsubsection{Precision}
To compare the precision of the two tools we employ the following precision metrics~\cite{elipmoc}:

\begin{itemize}[leftmargin=120pt]
  \item[Unresolved Operand:] Missing operands in the output.
  \item[Unstructured Control Flow:] High-level control flow in the output that is
    not expressible using structured programming constructs (e.g., high-level loops
    or conditionals).
  \item[Polymorphic Jump Target:] (intra-procedural) Jump instructions
    with targets not uniquely resolved under the same context.
\end{itemize}

\begin{figure*}[ht!]
  \centering
  \begin{subfigure}[b]{0.4\textwidth}
      \centering
      \includegraphics[height=2in]{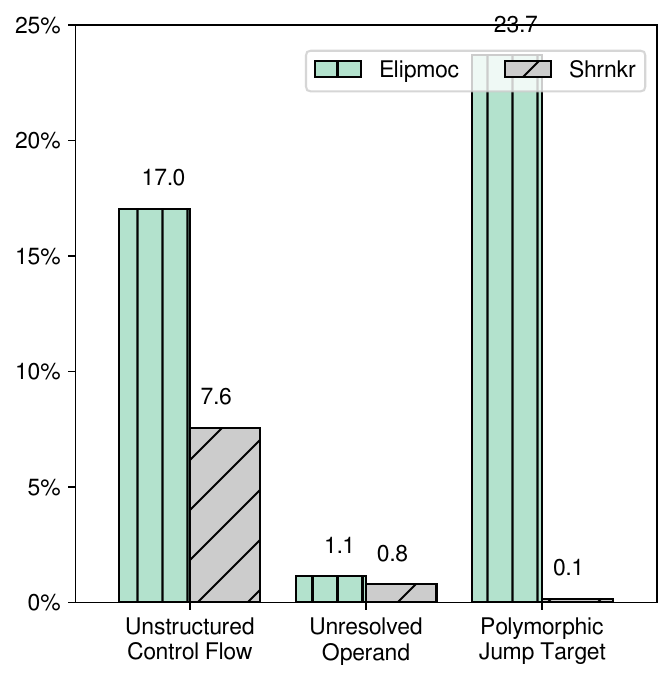}
      \caption{Standard Dataset}
  \end{subfigure}%
  ~
  \begin{subfigure}[b]{0.4\textwidth}
      \centering
      \includegraphics[height=2in]{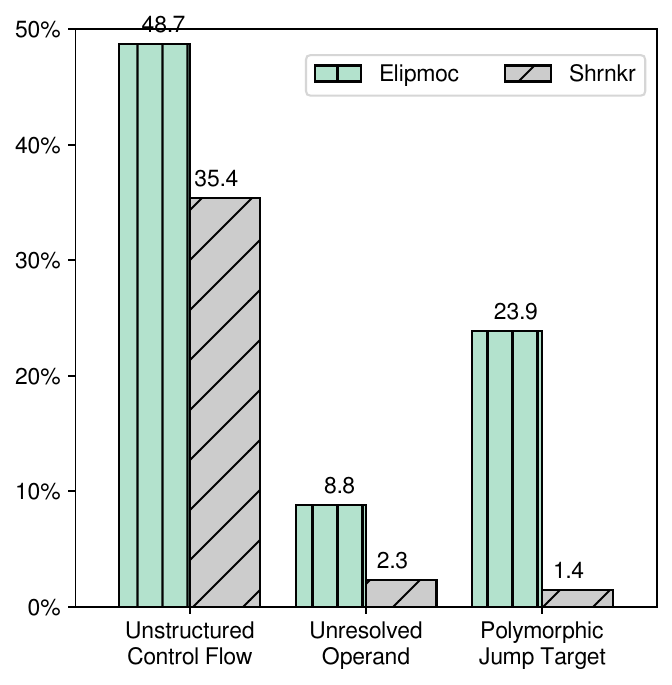}
      \caption{Yul Dataset}
  \end{subfigure}
  \caption{Precision Metrics in comparison with Elipmoc. All metrics
  show the \% of contracts over the common contracts the 2 tools manage
  to decompile that exhibit the behavior measured---lower is better.}
  \label{fig:elipmoc:precision}
\end{figure*}

The percentages of contracts which exhibit these imprecision artifacts for the subset of contracts
analyzed by both \ourtool{} and Elipmoc are available in Figure~\ref{fig:elipmoc:precision}.
For all metrics, \ourtool{} presents a clear improvement over Elipmoc.
Inspecting the results of the ``Polymorphic Jump Target'' metric one can clearly notice that imprecision
of the global control-flow-graph analysis has been nearly eliminated with 0.1\% of the contracts having some
imprecision compared to Elipmoc's 23.7\% for the Standard dataset, with 1.4\% and 23.9\% respectively for the
Yul dataset. 

Notably, Figure~\ref{fig:elipmoc:precision} \emph{downplays} the precision impact. If one takes
the improvement in the cumulative value of each metric (instead of just the percentage
of contacts that exhibit any non-zero amount of imprecision) the effect is magnified.
Table \ref{tab:elipmoc:metrics} presents these absolute numbers.
For example, for the ``Unstructured Control Flow'' metric on the Yul dataset, Figure~\ref{fig:elipmoc:precision}
shows a 27\% decrease in the number of contracts with imprecision (48.7\% to 35.4\%).
However, considering the absolute numbers, the total reduction of imprecision instances
is nearly 59\% (7,410 to 3,057).

\begin{table}[!htb]
  \caption{Analysis metrics for a comparison of \ourtool{} and Elipmoc. The table unifies both precision and completeness metrics.
  \textbf{Top table}: Standard Dataset. \textbf{Bottom table}: Yul Dataset.}
  \begin{minipage}{1\linewidth}
  \centering
  \begin{small}
  \begin{tabular}{|c|c|c|c|c|c|}
    \hline
    {} &  Polymorphic &  Missing &  Missing &  Unresolved &  Unstructured \\
    {} &  Jump Target &  Control Flow &  IR Block &  Operand &  Control Flow \\
    \hline
    Elipmoc &                     4118 &                  9411 &               1712 &                 202 &                       2253 \\
    \ourtool{}  &                       19 &                   424 &                  3 &                  108 &                        667 \\
    \hline
    \end{tabular}
  \end{small}
\end{minipage}

\vspace{0.7cm}
\begin{minipage}{1\linewidth}
  \centering
  \begin{small}
    \begin{tabular}{|c|c|c|c|c|c|}
      \hline
      {} &  Polymorphic &  Missing &  Missing &  Unresolved &  Unstructured \\
      {} &  Jump Target &  Control Flow &  IR Block &  Operand &  Control Flow \\
      \hline
      Elipmoc &                     2288 &                  1221 &               2443 &                2116 &                       7410 \\
      \ourtool{}  &                       74 &                   145 &                1161 &                 196 &                       3057 \\
      \hline
      \end{tabular}
  \end{small}
\end{minipage}
\label{tab:elipmoc:metrics}
\end{table}

\subsubsection{Completeness}

Static-analysis-based decompilers are nominally complete, i.e., cover all code. However,
this is not a full guarantee, for two reasons. First, the decompiler will likely have a bound in
the amount of work it performs, in order to minimize timeouts. Second, although each
statement may be covered, not all execution paths may be covered.

To compare the completeness of the two tools, we use two
incompleteness metrics:
\begin{itemize}[leftmargin=100pt]
  \item[Missing IR Block:] Blocks that are reachable in the global CFG analysis but
  do not have any corresponding blocks in the three-address IR (TAC) output.
  \item[Missing Control Flow:] Blocks in the TAC output that do not have the required number
  of outgoing edges (1 for non-return blocks, 2 for conditional jumps).
\end{itemize}

Both of these kinds of incompleteness artifacts arise due to the decompiler's inability to process
the input context-sensitive control-flow graph, to produce a normalized decompilation output.

\begin{figure*}[t!]
  \centering
  \begin{subfigure}[b]{0.4\textwidth}
      \centering
      \includegraphics[height=2in]{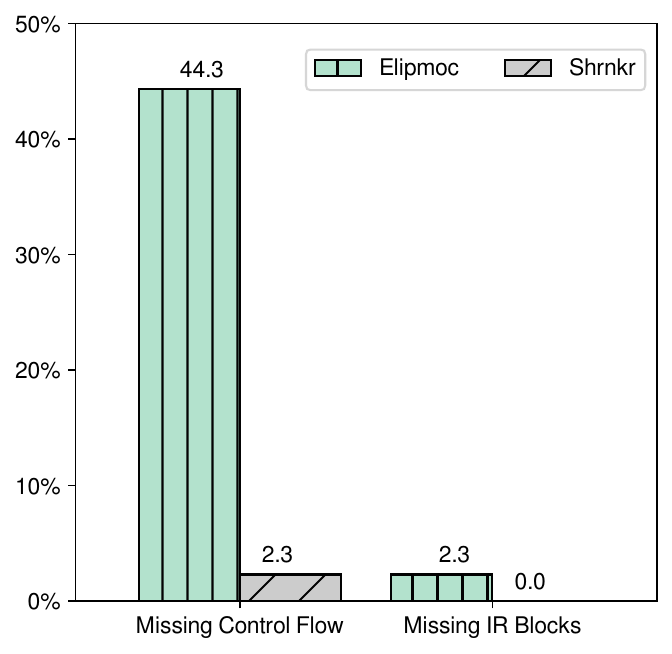}
      \caption{Standard Dataset}
  \end{subfigure}%
  ~
  \begin{subfigure}[b]{0.4\textwidth}
      \centering
      \includegraphics[height=2in]{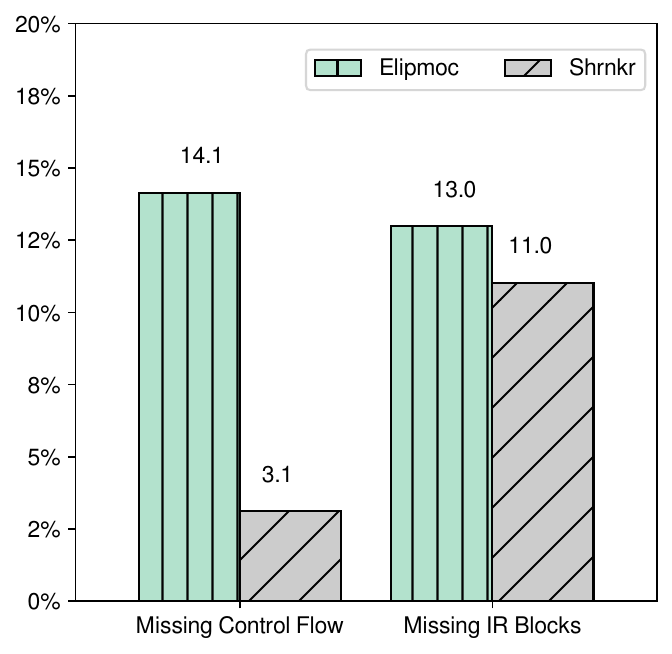}
      \caption{Yul Dataset}
  \end{subfigure}
  \caption{Completeness Metrics in comparison with Elipmoc. All metrics
  show the \% of contracts over the common contracts the 2 tools manage
  to decompile that exhibit the behavior measured---lower is better.}
  \label{fig:elipmoc:completeness}
\end{figure*}

The percentages of contracts that exhibit these incompleteness artifacts are plotted
in Figure \ref{fig:elipmoc:completeness} and the absolute counts are shown in Table~\ref{tab:elipmoc:metrics}.

As can be seen, \ourtool{} significantly lowers incompleteness. The only metric that still
exhibits non-negligible incompleteness artifacts is ``Missing IR Blocks'' and, although
11\% of decompiled contracts in the Yul dataset have at least one such block, the absolute
number of such missing blocks is tiny: just 0.11\% of total recovered basic blocks.

\subsection{Comparison with \heimdall{}}

\heimdall{}~\cite{heimdall} is an increasingly-popular symbolic-execution-based decompiler.
It has received significant attention in the past year, and its GitHub repository has surpassed 100 forks and 1,000 stars in
a brief time. The primary objective of \heimdall{} is to serve as a precise and performant decompilation toolkit.

With symbolic execution being the backbone of \heimdall{}, the decompilation leverages the symbolic
representation of a program's execution traces in order to produce a higher-level program representation.
This approach enables the decompiler to reason over actual execution paths, resulting in highly-precise
decompilation results. However, since it is only feasible to utilize a limited number of execution sequences,
symbolic-execution-based methods typically yield incomplete results, capturing fewer program behaviors overall.

Due to the fundamentally different architecture of \ourtool{} and \heimdall{}, we cannot directly
compare detailed internal metrics for each tool, as in the comparison with Elipmoc.
However, we can compare user-level, end-to-end metrics. Specifically, we compare the number of
\emph{unique} \textsc{CALL} and \textsc{LOGx} \emph{signatures} in the decompiled code. These are
code elements that should undisputedly exist in a correct decompilation: they are the static \emph{signatures} of functions called
on \emph{external} contracts (encoded in the bytecode as the 4-byte hash of the function name and argument types---e.g., \sv{0x0001e862} for ``\sv{balanceOfAt(uint256,address)}'') and of events emitted for consumption by external,
off-chain code (encoded as a 32-byte hash of a similar signature). Capturing (in decompiled code) these unique signatures
is a completeness/coverage metric over possible contract behaviors with regards to external
calls and events. Although simple, the metric has the property of being indifferent
to different decompilation styles (esp. inference of private functions by \ourtool{}
vs. inlining of all code/logic by \heimdall{}).

Table~\ref{tab:heimdal} shows the number of \textsc{CALL} and \textsc{LOGx} signatures
that are identified by each tool. (In these completeness numbers, \textbf{higher is better}.)
\ourtool{} manages to discover 67\% more calls (13,998 calls compared to \heimdall{}'s 8,381) in the standard dataset and
38\% more calls (13,600 calls against 9,841 for \heimdall{}) in the Yul dataset. Since the numbers provide an estimate of how much more code is 
decompiled by \ourtool{} when compared against \heimdall{}, the results  demonstrate the large advantage of \ourtool{} over \heimdall{} in 
terms of completeness. A similar conclusion may be drawn by looking at the events metrics.

\begin{table}[!htb]
  \caption{Total number of identified \textsc{CALL} and \textsc{LOGx} signatures between
  \label{tab:heimdal}
  \ourtool{} and \heimdall{}.\\
  \textbf{Top table}: Standard Dataset. \textbf{Bottom table}: Yul Dataset.}
  \begin{minipage}{0.8\linewidth}
  \begin{small}
    \begin{tabular}{|c|c|c|c|c|}
    \hline
    {}           &  Unique External Calls   &  Unique Events  &  Avg. Time & Timeouts \\
    \hline
    \ourtool{}   &                    13998 &         12725 &  1.87s       & 13        \\
    \heimdall{}  &                     8381 &           9345 &  0.88s      & 0         \\
    \hline
\end{tabular}

  \end{small}
  \end{minipage}

  \vspace{0.7cm}
  \begin{minipage}{0.8\linewidth}
  \begin{small}
    \begin{tabular}{|c|c|c|c|c|}
    \hline
    {}           &  Unique External Calls   &  Unique Events  &  Avg. Time & Timeouts \\
    \hline
    \ourtool{}   &                    13600 &         13661 &  6.76s       & 94        \\
    \heimdall{}  &                     9841 &           9505 &  0.59s      & 0         \\
    \hline
\end{tabular}

  \end{small}
  \end{minipage}
\end{table}

Table~\ref{tab:heimdall:metricssize} breaks down these results by contract size. As can be seen, the completeness benefit
is very substantial in large contracts, leading nearly to a doubling of event and function signatures observed in the output
code. It is reasonable to expect that larger contracts have a higher need for automatic analysis: they are both harder
to analyze manually and involve more sophisticated code patterns. Therefore, any verifiable advantage in completeness
holds large practical value. 

\begin{table}[!htb]
  \caption{\ourtool{} and \heimdall{}'s sigs for each contract size class. The number of contracts per size class is
    slightly smaller than in Table~\ref{tab:elipmoc:scale:depth} because timeouts are excluded.\\
  \textbf{Top table}: Standard Dataset. \textbf{Bottom table}: Yul Dataset}
\begin{minipage}{\linewidth}
  \centering
\begin{small}
\begin{tabular}{r|c|c|c|c|c}
    Bytecode Size & [0,5KB) & [5KB,10KB) & [10KB,15KB) & [15KB,20KB) & [20KB,max) \\
    \hline
  \ourtool{} function sigs   & 2343\textbf{(+37\%)} & 3245\textbf{(+58\%)}  & 3127\textbf{(+89\%)} & 2259\textbf{(+65\%)} & 3024\textbf{(+88\%)}  \\
  \heimdall{} function sigs  & 1699 & 2051  & 1654 & 1371 & 1606 \\
  \hline
  \ourtool{} event sigs   & 1995\textbf{(+15\%)} & 3050\textbf{(+19\%)}  & 2629\textbf{(+31\%)} & 1479\textbf{(+36\%)} & 3572\textbf{(+80\%)}  \\
  \heimdall{} event sigs  & 1729 & 2554  & 1994 & 1084 & 1984 \\
  \hline
  Contracts in size class     & 2550          & 1061         & 647          & 294          & 435
\end{tabular}
\end{small}
\end{minipage}

\vspace{0.7cm}
  \begin{minipage}{\linewidth}
  \centering
\begin{small}
  \begin{tabular}{r|c|c|c|c|c}
  Bytecode Size & [0,5KB) & [5KB,10KB) & [10KB,15KB) & [15KB,20KB) & [20KB,max) \\
  \hline
  \ourtool{} function sigs   & 2254\textbf{(+19\%)} & 3197\textbf{(+33\%)}  & 2384\textbf{(+39\%)} & 2296\textbf{(+40\%)} & 3469\textbf{(+57\%)}  \\
  \heimdall{} function sigs  & 1883 & 2397  & 1712 & 1643 & 2206 \\
  \hline
  \ourtool{} event sigs   & 1451\textbf{(+35\%)} & 3078\textbf{(+27\%)}  & 2727\textbf{(+31\%)} & 2340\textbf{(+57\%)} & 4065\textbf{(+65\%)}  \\
  \heimdall{} event sigs  & 1068 & 2410  & 2077 & 1487 & 2463 \\
  \hline
  Contracts in size class     & 1163          & 709         & 404          & 305          & 325
\end{tabular}
\end{small}
\end{minipage}
\label{tab:heimdall:metricssize}
\end{table}

Table~\ref{tab:heimdal} also shows average execution time and repeats the \ourtool{} timeout rate.
It is apparent that, in terms of scalability, \heimdall{} has no hurdle to overcome, as expected in a symbolic
execution tool, which covers the program only to the extent that it can execute it precisely. \heimdall{} has
no timeouts and is extremely fast on average. The \heimdall{} average execution time can be more than 10 times
smaller than \ourtool{}---although the average times
are low for both tools, at under 7s for the slowest dataset.

Overall, the results are indicative of the completeness superiority of a static analysis when compared against symbolic execution as the underlying technique for decompilation. Arguably, the very essence of a decompiler is to lift as much low-level code as possible.
Thus, \textit{every} part of the program to be decompiled should be considered, and while symbolic execution makes an efficient implementation simpler from an 
engineering standpoint, the completeness offered by a deep static analysis appears to be unparalleled.

\subsection{Human Study}
\label{sec:study}

To provide deeper insights in our comparison with Elipmoc and \heimdall{}, we complement the quantitative evaluation
of the two previous subsections with a small-scale human study, assessing the decompilation quality of the outputs
of the 3 tools.

To produce source-like high-level output (instead of the usual TAC output of \ourtool{}) we ported the source unparser of the Dedaub Security
Suite to use \ourtool{}.

We directed the study to expert participants working in the industry and/or academia in roles related to smart contract
security. To incentivize participation, we offered participants a \$100 reward.
Overall, 8 participants completed the study, each given 3 randomly assigned decompilation tasks, resulting in a total of 24 data points.

\begin{wrapfigure}{l}{0.5\textwidth}
  \includegraphics[height=2.65in,clip,trim=12px 0px 0 0px]{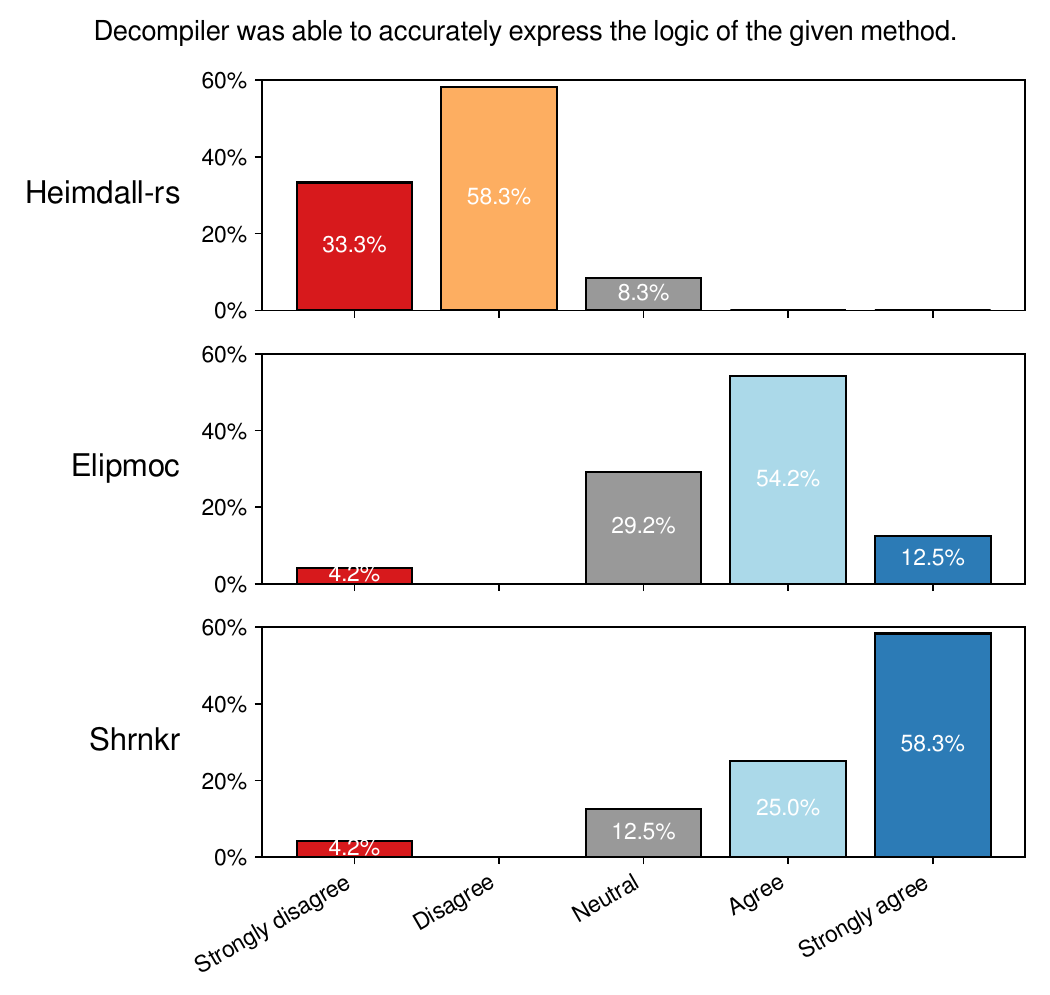}
  \caption{Human Study Results: Participant agreement with statement \textit{"Decompiler was able to to accurately express the logic of the given method."}}
  \label{fig:study}
\end{wrapfigure}

More specifically, for each decompilation task the participant was given the outputs of the three tools as well as
the original source code and was tasked with evaluating the decompilers' ability to accurately recover the logic
of the contract's largest public method (which was identified programmatically and provided to the participants).

To minimize participant bias we anonymized the outputs of the three tools and prompted the participants to \textit{``attempt to ignore the differences in decompilation
style such as the inference of private functions vs the inlining of all code, naming conventions, etc.''}.
The entirety of the study is available in \AppendixC{}.

Figure \ref{fig:study} plots the results of the human study.
\heimdall{} is shown to perform the worst in the human study, with its incomplete algorithms failing to recover the logic of
the given programs. While both \ourtool{} and Elipmoc have a generally positive performance with the majority of participants
rating their decompilation ability positively, \ourtool{}'s improvements manage to shift the participant concensus from "Agree"
to "Strongly Agree", highlighting their usefulness.

\subsection{Case Study: Decompilation of Hack Contracts}

As a case study, we examine the ability of \ourtool{}, Elipmoc, and \heimdall{} to decompile adversarial contracts,
used in past security attacks (``hacks''). The goal is to confirm that there is no obvious negative bias in this
subset of contracts, at least with respect to scalability.
The case study is performed on the \sv{malicious\_smart\_contracts} dataset
\footnote{\url{https://github.com/forta-network/labelled-datasets/blob/main/labels/1/malicious_smart_contracts.csv}}
from the labelled-datasets repository of OpenZeppelin's Forta Network.
While the original dataset contains 753 smart contract addresses, it contains many duplicate bytecodes which we removed,
ending up with a dataset of 592 unique contracts.


\begin{wraptable}{l}{0.4\textwidth}
\caption{Runtime statistics for the adversarial dataset.}
\label{tab:hack:scale}
\begin{small}
\begin{tabular}{|c|c|c|}
  \hline
  {}          & Avg. Time & Timeouts \\
  \hline
  \ourtool{}  &     0.79s &   0  \\
  Elipmoc     &     0.67s &  10 (\textbf{1.69\%}) \\
  \heimdall{} &     0.44s &   0 \\
  \hline  
\end{tabular}
\end{small}
\end{wraptable}
Table \ref{tab:hack:scale} presents the runtime statistics for the adversarial contracts dataset.
As can be seen, these contracts represent a smaller challenge than our other two datasets, with \ourtool{} and \heimdall{} managing to decompile
all 592 contracts and Elipmoc timing out at only 1.69\% percent of contracts. This result can be explained by the relative
simplicity of hack contracts: regular contracts, intended to be used heavily, pack as much code as possible and employ
heavy optimization, whereas adversarial (hack)
contracts are both smaller (i.e., do not struggle to fit in the EVM 24KB limit) and less mature.
Accordingly, in terms of decompilation completeness, in the comparison between \ourtool{} and \heimdall{}, the story of Table \ref{tab:heimdal}
stands: \ourtool{} recovers 60.5\% more external call signatures (1106 vs 689) and 9.4\% more event signatures (1049 vs 959).

\subsection{Design Decisions}

In order to understand how each of the features of \ourtool{} affect its scalability,
precision, and completeness, we decompiled our datasets using 3 modified configurations of \ourtool{},
in addition to its default one:

\begin{itemize}[leftmargin=80pt]
  \item[No Shrinking ctx:] \ourtool{} replacing its \emph{shrinking context sensitivity} with the
  \emph{transactional context sensitivity} of Elipmoc.
  \item[No Cloning:] \ourtool{} with the cloning transformation we presented in
    Section \ref{sec:cloning} disabled.
  \item[No Pre-Analysis:] \ourtool{} with the incomplete global pre-analysis
  we presented in Section \ref{sec:incompleteGlobal} disabled.
\end{itemize}

\myparagraph{Scalability} Table
\ref{tab:decisions:scale} shows the timeouts for the various different
configurations of \ourtool{}. (In the metrics of this section, \textbf{lower is better}.)

\begin{table}[!htb]
  \caption{Timeouts for \ourtool{} configurations.
    \textbf{Left table}: Standard Dataset. \textbf{Right table}: Yul Dataset.}
  \begin{small}
  \begin{minipage}{.5\linewidth}
    \centering
    \begin{tabular}{c|c|}
      {} &  Timeouts \\
      \hline
      No Shrinking ctx &      559 (\textbf{11.18\%}) \\
      No Cloning       &       16 (\textbf{0.32\%}) \\
      No Pre-Analysis    &       19 (\textbf{0.38\%}) \\
      \ourtool{}           &       13 (\textbf{0.26\%}) \\
      \hline
      Total contracts in dataset &     5000 \\
      \end{tabular}      
  \end{minipage}%
  \begin{minipage}{.5\linewidth}
    \centering
      \begin{tabular}{|c|c}
        {} &  Timeouts \\
        \hline
        No Shrinking ctx &      403 (\textbf{13.43\%}) \\
        No Cloning       &      105 (\textbf{3.5\%}) \\
        No Pre-Analysis    &      146 (\textbf{4.86\%}) \\
        \ourtool{}           &       94 (\textbf{3.13\%}) \\
        \hline
        Total contracts in dataset  &     3000 \\
        \end{tabular}
        
  \end{minipage}
  \end{small}
  \label{tab:decisions:scale}
\end{table}

The table makes clear that the scalability of \ourtool{} is due to the \emph{shrinking context sensitivity}.
Disabling it leads to 35x the timeouts for the Standard dataset and over 4x the timeouts on the Yul dataset.
In addition, the more demanding Yul dataset allows us to observe that all 3 of the components
of \ourtool{} have a positive impact on scalability. Disabling the incomplete global pre-analysis
leads to a 53\% increase in timeouts while disabling the cloning transformation leads to
a 10\% increase.

\begin{table}[!tb]
  \caption{Analysis metrics for various configurations of \ourtool{}. The table unifies both precision and completeness metrics.
    \textbf{Top table}: Standard Dataset. \textbf{Bottom table}: Yul Dataset}
  \begin{minipage}{1\linewidth}
  \centering
  \begin{small}
    \begin{tabular}{|c|c|c|c|c|c|}
    \hline
    {} &  Polymorphic &  Missing &  Missing &  Unresolved &  Unstructured \\
    {} &  Jump Target &  Control Flow &  IR Block &  Operand &  Control Flow \\
    \hline
    No Shrinking ctx &                     3406 &                   459 &                751 &                 128 &                        611 \\
    No Cloning       &                       18 &                   728 &                153 &                 140 &                       1042 \\
    No Pre-Analysis    &                       40 &                   421 &                  3 &                  62 &                        542 \\
    \ourtool{}           &                       19 &                   421 &                  3 &                  62 &                        542 \\
    \hline
    \end{tabular}    
  \end{small}
\end{minipage}

\vspace{0.7cm}
\begin{minipage}{1\linewidth}
  \centering
  \begin{small}
    \begin{tabular}{|c|c|c|c|c|c|}
      \hline
      {} &  Polymorphic &  Missing &  Missing &  Unresolved &  Unstructured \\
      {} &  Jump Target &  Control Flow &  IR Block &  Operand &  Control Flow \\
      \hline
      No Shrinking ctx &                     3381 &                   253 &               1474 &                 333 &                       3624 \\
      No Cloning       &                       34 &                   168 &               1602 &                 504 &                       6568 \\
      No Pre-Analysis    &                      101 &                   121 &                956 &                 184 &                       2927 \\
      \ourtool{}           &                       65 &                   120 &                956 &                 184 &                       2927 \\
      \hline
      \end{tabular}
  \end{small}
      
\end{minipage}
\label{tab:decisions:metrics}
\end{table}

\myparagraph{Precision and Completeness}
Table \ref{tab:decisions:metrics} presents the absolute numbers for our precision and completeness metrics for the
four configurations of \ourtool{}.

It is easy to see that the replacement of the \emph{shrinking context sensitivity} with Elipmoc's
\emph{transactional context sensitivity} produces a much less precise analysis.
This imprecision first manifests itself at the global context-sensitive control-flow graph level
and also results in decompilation artifacts at the three-address-code output.

Disabling the cloning component has the biggest negative impact to the precision of \ourtool{}'s decompilation
output with 92\% more ``Unstructured Control Flow'' inferences for the Standard dataset and 124\% for the Yul dataset.

Lastly, disabling the incomplete global pre-analysis results in imprecision
for the global control-flow graph analysis, without affecting the precision of three-address-code output.

Inspecting the completeness metrics, we can deduce that the \emph{shrinking context sensitivity} and block cloning techniques
have the largest impact on the completeness of \ourtool{}.

\section{Related Work}
\label{sec:related}

Multiple EVM decompilers have been proposed over the years \cite{Suiche2017,gigahorse,panoramix,vandal}.
However, continuous technical advancements are needed to keep up with the complexity of modern smart contracts, therefore
many decompilers (even past leaders) have been not been maintained \cite{Suiche2017,panoramix,vandal},
delivering very poor results by more modern standards \cite{gigahorse,elipmoc}.
Most relevant to our work, Gigahorse~\cite{gigahorse}, Elipmoc~\cite{elipmoc}, and \heimdall{} have been discussed extensively throughout the paper.
Other decompilers used by practitioners include EtherVM \cite{ethervm}, and, indirectly, the decompiler in Certora \cite{sagiv2020invited,CertoraMemory}.
More recent decompilers used for static analysis clients include Ethersolve~\cite{ethersolve}.
However, \cite{ethersolve} only raises the abstraction level to a global CFG, which requires a small subset of the techniques developed within \ourtool{}.
For instance, Ethersolve does not produce a register-based IR (retaining the original stack-altering instructions in its output)
nor does it discover private functions. These limitations inhibit its ability to support high-level client analyses.

A number of other EVM toolchains are used today, and several studies~\cite{ChaliasosStudyICSE24,StudyICSE23,PerezStudyUSENIX21} have examined their usefulness and real-world impact.
Among them, popular for finding vulnerabilities, are fuzzing frameworks~\cite{ContractFuzzer,harvey,SMARTIAN,LearningtoFuzz}, which identify vulnerabilities by analyzing bytecode directly.
Notable examples include ContractFuzzer~\cite{ContractFuzzer}, Harvey~\cite{harvey}, Echidna~\cite{echidna,echidna21}, sFuzz~\cite{SFuzz},
and the recent Ityfuzz~\cite{ityfuzz}, which leverages a faster interpreter (RETH) for improved performance.
Additionally, are a number of tools, which are meant to analyze the 3-address
IR output that \ourtool{} provides, including MadMax~\cite{MadMax}, Ethainter~\cite{ethainter}, Greed~\cite{ConfusumContractum,NotYourType}, DeepInfer~\cite{DeepInfer}, and Todler \cite{TODLER}.

Outside of the smart contract domain, a number of tools and techniques are relevant.
Context sensitivity has been employed in many static analysis settings before, and is well-known for improving precision for value-flow analysis
in languages with dynamic dispatch \cite{dvanhorn:milanova-etal-tosem05,10.1145/1925844.1926390, 10.1145/3540250.3549097,dvanhorn:Sharir:Interprocedural}.
Selective context sensitivity approaches~\cite{UnityRelay,Zipper,Introspective,SelectiveImpactPre,kLimiting,DataDriven2018,DataDriven2019,ReturnofCFA,DataDriven2017,eagle} have been proposed to 
overcome the scalability and precision obstacles faced when applying traditional context sensitivity
\cite{dvanhorn:milanova-etal-tosem05,10.1145/1925844.1926390,Shivers1991} variants to large, real-world programs. Much past selective context sensitivity research \cite{Introspective,SelectiveImpactPre,Zipper,ZipperE,Scaler} has relied on the
results of a pre-analysis to create context sensitivity variants that achieve balance between scalability and precision.
Such work \cite{Introspective,Zipper,ZipperE,Scaler,kLimiting} often makes use of an imprecise context-insensitive pre-analysis, which is
not always ideal when attempting to approximate the behavior of a context-sensitive analysis. \emph{Shrinking context sensitivity} does not
base its decisions on such a less-precise pre-analysis.

In some of the aforementioned work \cite{DataDriven2017,DataDriven2018,DataDriven2019,ReturnofCFA}, selective context sensitivity has also been fruitfully combined with machine learning approaches. In \cite{DataDriven2018}
authors introduced the technique of \emph{context tunneling} to create context sensitivity variants that, upon a transition, in some cases choose to update
the calling context and in others to simply propagate it. Context tunneling has shown great promise in the analysis of Java applications, having been used~\cite{ReturnofCFA}
to almost completely simulate object sensitivity via call-site sensitivity.
Our \emph{shrinking context sensitivity} (as well as Elipmoc's \emph{transactional context sensitivity}) also employs similar logic to just propagate (instead of updating) the calling
contexts in most transitions.

An important distinction relative to \ourtool{} is that all such past work (in selective context sensitivity) limits the context by \emph{avoiding} to include context elements, in advance. For instance, the description
of novelty of the \textsc{Bean} technique \cite{kLimiting} reads: 
\emph{The novelty lies in identifying redundant context elements [...] based on a pre-analysis (e.g., a context-insensitive Andersen’s analysis) performed initially on a program, and then avoid them in the subsequent k-object-sensitive analysis.} All tunneling work follows a similar pattern.

Instead, the distinguishing feature of \emph{shrinking context sensitivity} is that it has a temporal character: it first includes context elements, while they are useful, but later eliminates them eagerly, i.e., before the maximum context depth is reached. No past algorithm has this feature.

It is not entirely surprising that past context-sensitive algorithms have not explored this direction. Past context-sensitive analyses have mainly worked in the setting of points-to analysis of large Java programs. The context depth employed in such a setting is much shorter and does not lend itself to more adaptive algorithms. For instance, the typical context depth in points-to analysis work~\cite{UnityRelay,Zipper,Introspective,SelectiveImpactPre,kLimiting} is just 2. Shrinking context sensitivity is applied with contexts of depth around 20.
This showcases that decompilation is a much higher-precision setting (but for very specific kinds of information).
This large context depth is a key enabler of shrinking context sensitivity: it means the context includes elements all the way from a function's call to its return, even if the function itself makes many other nested calls.

Binary disassembly~\cite{SpeculativeDisassembly,floresmontoya2019datalog,StaticDisassembly}
and decompilation~\cite{180374,quteprints36820,nomoregotos,Emmerik2007StaticSA,8330222,HelpingJohnny} techniques have seen
use in several domains.
Numerous foundational techniques had been established by the mid-1990s~\cite{quteprints36820},
with particular emphasis on the x86 architecture. This architecture offers a somewhat simplified path to decompilation,
given a dependable disassembly process. The delineation of function boundaries and the deduction of arguments are facilitated
by the adherence to standard calling conventions, the Instruction Set Architecture's (ISA) support for function calls and returns,
and a uniform call stack architecture. More closely aligned with the techniques of our work,
the Ddisasm tool~\cite{floresmontoya2019datalog} uses Datalog to provide a disassembler for x64 binaries,
while the OOAnalyzer system~\cite{PrologRecover} employs a logic programming approach with Prolog to infer C++ class structures from compiled binaries.

\section{Conclusion}

We presented \ourtool{}, a static-analysis-based decompiler for EVM bytecode.
\ourtool{} achieves a significant improvement over the state-of-the-art using a new
variant of context sensitivity, \emph{shrinking context sensitivity}, additionally tuned via an incomplete
global pre-analysis, and a cloning transformation to better normalize decompilation output.
These three advancements enable \ourtool{} to vastly outscale the state-of-the-art
decompiler, while also seeing significant improvements in both precision and completeness.
\ourtool{} was also compared against the most popular alternative-technology decompiler
displaying superior coverage of program behaviors.
We perform our evaluation on datasets of contracts using the two pipelines of the
Solidity compiler: the currently default ``legacy'' pipeline, and the new Yul pipeline.
The latter had not been considered in the evaluations of previous publications, and we
experimentally show it to provide a greater challenge to decompilers.

\section*{Acknowledgments}

We thank the study participants for taking the time to complete the study.
We gratefully acknowledge funding by ERC Advanced Grant \textsc{PINDESYM} (101095951).

\section*{Data-Availability Statement}

\ourtool{} is part of the \href{https://github.com/nevillegrech/gigahorse-toolchain/tree/sub24}{gigahorse-toolchain} open source repository.
The paper's artifact is also publicly available ~\cite{ShrnkrArtifact} and can be used to reproduce all experiments in the paper's evaluation
except for the human study in Section~\ref{sec:study}, which uses closed-source code.

\bibliography{bibliography,references,tools}

\newpage
\appendix

\section{Control Flow Normalization via Cloning}\label{sec:cloningFull}

A second technique that enables precision in \ourtool{} is the aggressive cloning of blocks that are \emph{locally}
determined to be used in inconsistent ways. (\emph{Local} inspection refers to inspection that does not require
the full power of the decompiler's static analysis, i.e., the shrinking context of Section~\ref{sec:shrinking}.
Effectively, the inspection examines the block's contents and occurrences of the block address constant in the contract
code.)

\subsection{Cloning Need: Illustration}
As mentioned in Section~\ref{sec:background}, the Solidity compiler will often try to reuse the same low-level blocks for different high-level purposes.
Such an optimization is essential in the setting of smart contracts as the Ethereum blockchain enforces a size limit of 24576 bytes~\cite{eip-170}.

The code snippet of the earlier Figure \ref{fig:calls:chained} exhibits such behavior with block \sv{0x72},
which is pushed to the stack twice as a continuation, in the course of performing two of the three checked subtraction
operations (per line 5 of our earlier Solidity example of Figure~\ref{fig:solidity-example}).

For this example, the state-of-the-art Elipmoc decompiler will correctly identify that this block is used to perform two different function calls, attempting to summarize them as
shown in the three-address-code snippet in Figure \ref{fig:chained:elipmoc}.

\begin{figure}[]
  \centering
  \vspace{-0.2cm}
  \begin{subfigure}[b]{0.95\textwidth}
    \includegraphics[width=\linewidth]{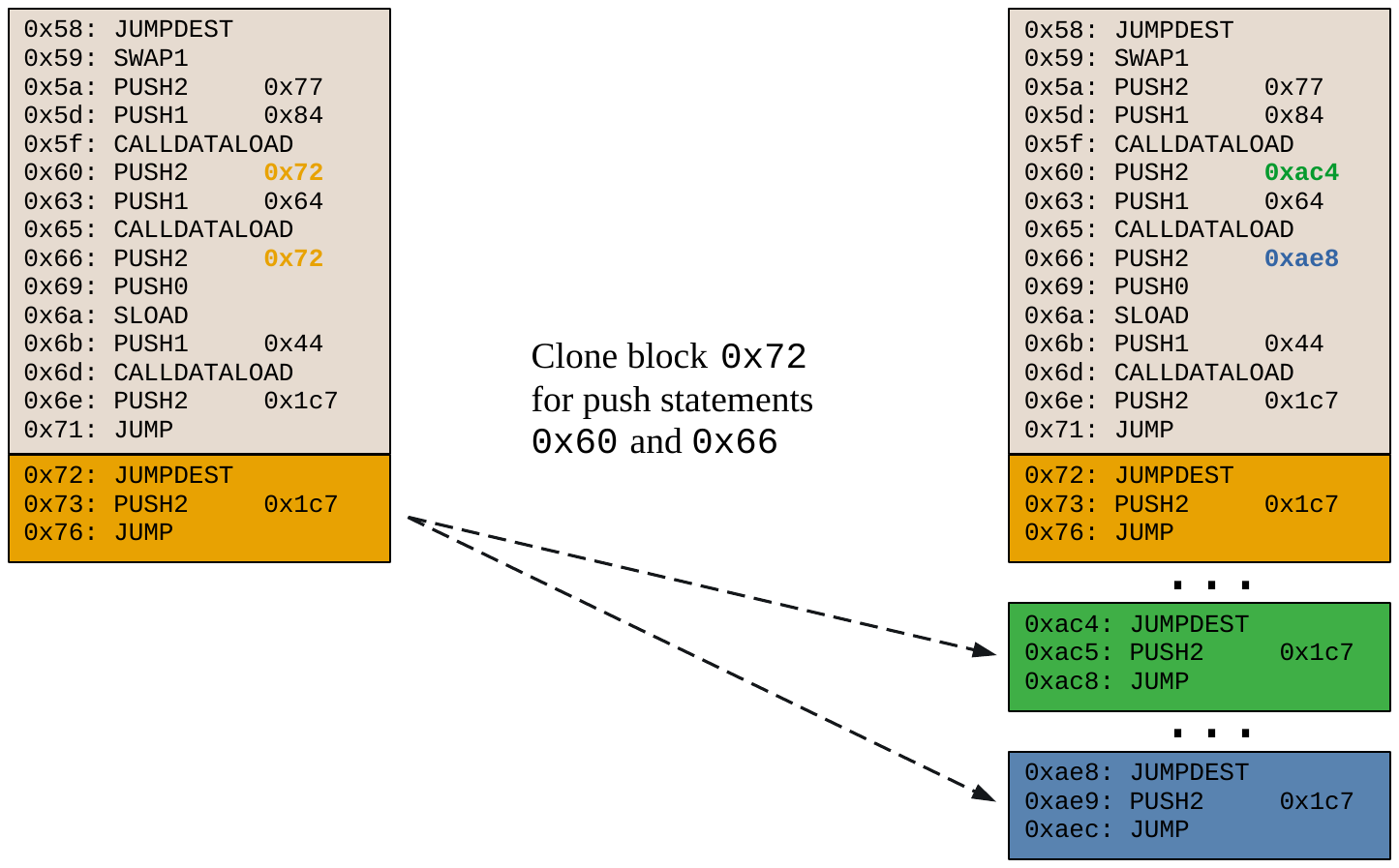}
    \vspace{-2.2cm}\\
    \caption{Cloning transformation: block-level view}
      \label{fig:chained:illustration}
  \end{subfigure}
  \begin{subfigure}[b]{0.5\textwidth}
      \centering
      \begin{EVMnonumbercodesmallwbox}
Begin block 0x58
prev=[0x178B0x50], succ=[0x72]
=================================
0x5a: v5a(0x77) = CONST
0x5d: v5d(0x84) = CONST
0x5f: v5f = CALLDATALOAD v5d(0x84)
0x60: v60(0x72) = CONST
0x63: v63(0x64) = CONST
0x65: v65 = CALLDATALOAD v63(0x64)
0x66: v66(0x72) = CONST
0x69: v69 = PUSH0
0x6a: v6a = SLOAD v69
0x6b: v6b(0x44) = CONST
0x6d: v6d = CALLDATALOAD v6b(0x44)
0x6e: v6e(0x1c7) = CONST
0x71: v71_0 =CALLPRIVATE v6e(0x1c7), v6d, v6a, v66(0x72)

Begin block 0x72
prev=[0x72, 0x58], succ=[0x72, 0x77]
=================================
0x72_0x0: v72_0 = PHI v71_0, v76_0
0x72_0x1: v72_1 = PHI v47(0x20), v5f, v65, v150V44
0x72_0x2: v72_2 = PHI v60(0x72), v44(0xab), v5a(0x77), v166V50
0x73: v73(0x1c7) = CONST
0x76: v76_0 = CALLPRIVATE v73(0x1c7), v72_0, v72_1, v72_2 // last argument is the continuation
        \end{EVMnonumbercodesmallwbox}
      \caption{Elipmoc Output}
      \label{fig:chained:elipmoc}
  \end{subfigure}%
  ~
  \begin{subfigure}[b]{0.5\textwidth}
      \centering
      \begin{EVMnonumbercodesmallwbox}
Begin block 0x58
prev=[0x50], succ=[0xae8]
=================================
0x5a: v5a(0x77) = CONST
0x5d: v5d(0x84) = CONST
0x5f: v5f = CALLDATALOAD v5d(0x84)
0x60: v60(0xac4) = CONST
0x63: v63(0x64) = CONST
0x65: v65 = CALLDATALOAD v63(0x64)
0x66: v66(0xae8) = CONST
0x69: v69(0x0) = CONST
0x6a: v6a = SLOAD v69(0x0)
0x6b: v6b(0x44) = CONST
0x6d: v6d = CALLDATALOAD v6b(0x44)
0x6e: v6e(0x1c7) = CONST
0x71: v71_0 = CALLPRIVATE v6e(0x1c7), v6d, v6a, v66(0xae8)

Begin block 0xae8
prev=[0x58], succ=[0xac4]
=================================
0xae9: vae9(0x1c7) = CONST
0xaec: vaec_0 = CALLPRIVATE vae9(0x1c7), v71_0, v65, v60(0xac4)

Begin block 0xac4
prev=[0xae8], succ=[0x77]
=================================
0xac5: vac5(0x1c7) = CONST
0xac8: vac8_0 = CALLPRIVATE vac5(0x1c7), vaec_0, v5f, v5a(0x77)
        \end{EVMnonumbercodesmallwbox}

      \caption{\ourtool{} Output}
      \label{fig:chained:shrinker}
  \end{subfigure}
  \caption{Cloning illustration and decompilation output of optimized chained call pattern using Elipmoc (i.e., without cloning support) and \ourtool{}}
\end{figure}

The results of this merging of different high-level calls are detrimental to the precision of downstream analyses.
Following the function call at block \sv{0x72}, the control-flow is transferred either to itself or to block \sv{0x77}.
(Note this in the \sv{succ} annotation in the block's header.)
The Elipmoc output in Figure \ref{fig:chained:elipmoc} simply cannot be expressed using standard linear IR control-flow constructs
over the basic blocks shown: the decompilation output needs to explicitly refer to passed continuations to explain the successor
blocks.

Such artifacts of imprecise decompilation output have been identified in past literature~\cite{elipmoc} as ``Unstructured Control Flow''.

In addition, the identities of the arguments of the two function calls are not recoverable.
Through the introduction of phi functions, the first argument (\sv{v72\_0}) is the merging of two variables
and the second one (\sv{v72\_1}) the merging of 4 variables.
As a result, an analysis that intends to precisely model the effects of the high-level calls will need to consider 8
combinations of arguments, 6 of which are invalid.

Both of the above problems are addressed in the output of \ourtool{} in Figure~\ref{fig:chained:shrinker}.
This is achieved via the low-level cloning transformation shown in \ref{fig:chained:illustration}:
For each of the two low-level statements pushing block \sv{0x72} to the stack, we generate a new cloned copy of
the block and replace their pushed value with the identifier of their corresponding fresh block.

To improve the precision and completeness of our decompilation output we employ heuristics to identify low-level blocks that are likely
reused to implement more than one high- or low-level construct and clone all their different uses.

\subsection{Identifying Block Cloning Candidates}

We will now describe the two block categories we identified as candidates for cloning.
It should be noted that both of these block categories describe small basic blocks with few to no
high-level operations in them.

\myparagraph{Reused Continuations}
The first class of blocks we identify as candidates for cloning are blocks that are used as continuations in
more than one case (i.e., by more than one push statement). These continuation blocks are often used
to perform chained calls at different points in a contract's execution (as in the example in Figure \ref{fig:calls:chained})

\myparagraph{Stack Balancing Blocks}
The second class of blocks we attempt to clone are \emph{Stack-Balancing Blocks}.
These are blocks that contain only low-level stack-altering instructions (i.e., \sv{POP}, \sv{SWAPx}, \sv{DUPx})
and end with a \sv{JUMP}.
An example \emph{Stack Balancing Block} can be seen in the following EVM code snippet:
\begin{EVMnonumbercode}
  0x1c3: JUMPDEST
  0x1c4: SWAP1
  0x1c5: POP
  0x1c6: SWAP2
  0x1c7: SWAP1
  0x1c8: POP
  0x1c9: JUMP
\end{EVMnonumbercode}

\emph{Stack Balancing Blocks} are \emph{locally-unresolved} blocks that are typically highly-reused in a single contract.
These blocks are often used to implement the return blocks of different private functions or even used in different ways
inside the same private function.

\subsection{Block Cloning Details}

After identifying the block-cloning candidates, we find the ones that are used in more than one place in the low-level bytecode
(i.e., are pushed to the stack by more than one bytecode statement).
We encode our block-cloning instances as tuples of [pushStmt, blockToClone] and generate a new low-level block for each tuple.
To reduce implementation complexity we only allow the cloning of low-level blocks that end with \sv{JUMP} statements,
having no fallthrough block that would need to be cloned as well.

\section{Incomplete Global Pre-Analysis}\label{sec:incompleteGlobalFull}

The general framework of shrinking context sensitivity admits several
refinements.
The last new technique from \ourtool{} that we present plays
multiple roles, e.g., in eliminating spurious call edges as well as in
introducing more edges than mere calls into the context.  Both cases
require a global pre-analysis, i.e., perform a best-effort,
\emph{incomplete} version of the full decompiler static analysis and
use it to configure the subsequent complete analysis.

\subsection{Overview}

The main context-sensitive global control-flow analysis is preceded by an \emph{incomplete} global pre-analysis round.
The pre-analysis helps in the following aspects of the decompiler.

\myparagraph{Spurious \predname{PrivateCallAndContinuation}}
Recall the \predname{PrivateCallAndContinuation} predicate we used when defining our context
sensitivity model. The default meaning of this predicate would be as-defined in prior work:
these \emph{likely} function calls are identified in the global analysis model of
the Gigahorse~\cite{gigahorse} decompiler, which is also used by the Elipmoc~\cite{elipmoc} decompiler.
Specifically:
\begin{bullets}
  \item A \emph{likely} function call is a block that jumps to another block directly, leaving
    the stack with one or more \emph{likely} continuations on it.
  \item A value pushed to stack is considered a \emph{likely} continuation if a \sv{JUMPDEST}
    exists at that index.
\end{bullets}

Thus if a block leaves the stack with a value that happens to be the same as a \sv{JUMPDEST} but it does not end up
being used as a jump target, our \textbf{Merge} constructor will add that \emph{spurious} caller to the calling context.
\ourtool{} refines this definition with a much deeper pre-analysis step.

\myparagraph{Spurious \predname{PublicCall}}
The definition of our \emph{shrinking context sensitivity} model also relies on the \predname{PublicCall} predicate,
locating the entries of the contract's public functions.
As the results of the \predname{PublicCall} are utilized by its global control-flow graph analysis,
Elipmoc uses local heuristic patterns to identify public function entries.

These patterns identify comparisons with small constant values (like \sv{0x12e49406} and \sv{0x87d7a5f4} in
our example in Figure \ref{fig:pubFunn}) that are \emph{likely} to be function selector values.
However one cannot confirm that these values are compared against the function selector bytes
loaded from the call-data using only a local analysis.
Such imprecise \predname{PublicCall} inferences can have a negative impact on our context-sensitive
control-flow graph.

\myparagraph{Imprecision-Introducing Edges}
The heuristics used in shrinking context sensitivity aim to approximate private function calls and returns.
However there can be non-call edges that would make our global analysis more precise if they were to be included in the calling
context. The decompiler can include a generic, agnostic (to control-flow idioms, such as ``calls'' and ``returns'')
way to identify such edges, based on blocks that introduce global
imprecision.

\subsection{Implementation}

The implementation of the incomplete global analysis is rather straightforward; we halt the execution of the global
analysis after a set number of facts have been produced.

\ourtool{} is implemented as a set of Datalog inferences, using the Souffle~\cite{Jordan16, DBLP:conf/cc/ScholzJSW16,souffleInterpreted} engine.
We restrict the global pre-analysis to a bounded amount of work by using souffle's \sv{.limitsize} directive
\footnote{\url{https://souffle-lang.github.io/directives\#limit-size-directive}} on the key \predname{BlockOutput} relation.
This directive will halt the bottom-up evaluation of predicate \predname{BlockOutput} once its size limit is reached. (We use a
coarse limit of 1,000,000 facts, a number derived empirically, but not particularly tuned.)
In addition, reaching the limit will also halt the evaluation of all other mutually-recursive relations in the same stratum, essentially stopping
the incomplete global pre-analysis, as we intend.

The inference logic of the incomplete global pre-analysis uses core relations in the
decompiler. These are summarized in Figure~\ref{fig:schema2} but are essentially a model that was inherited
from past static-analysis-based decompilers, tracing back to Gigahorse~\cite{gigahorse}.

\begin{figure*}[tb!p]
\begin{small}
  \begin{tabular}{l}
  $V$: set of virtual ``variables'', i.e., instructions that push values or stack positions at block entry  \\
  $B$: set of basic blocks \\
  $PC$: set of private contexts, \args{PC} $\cong$ $B^n$ \\
  $C$: set of contexts, \args{C} $\cong$ \args{B $\times$ PC}, as labeled record \record{\args{B}}{\args{PC}}\\
  \end{tabular}

\begin{tabular}{l l l}
\cline{1-2}
\pred{BlockInput}{ctx: C, block: B, i: $\mathbb{N}$, var: V} & \multirow{2}{*}{$\begin{cases} \shortstack[l]{At entry/exit to/from \args{block}, under context \args{ctx},\\
  the \args{i}-th position of the stack contains \args{v}.} \end{cases}$}\\
\pred{BlockOutput}{ctx: C, block: B, i: $\mathbb{N}$, var: V} &  \\
\pred{BlockJumpTarget}{ctx: C, from: B, var: V, to: B} &  Block \args{from}, under context \args{ctx}, pushes \args{var} on the \\
  &  stack, whose value is a jump target, \args{to}.\\
\pred{GlobalBlockEdge}{fromCtx: C, from: B, toCtx: C, to: B} & Block \args{from} under context \args{fromCtx}\\
  &  may jump to block \args{to} under context \args{toCtx}.\\
\cline{1-2}
\noalign{\vskip 1mm}
\end{tabular}
\end{small}
\caption[]{Core relations in decompiler, used in global pre-analysis.}
\label{fig:schema2}
\end{figure*}

\subsection{Removing Spurious PrivateCallAndContinuation facts}

The incomplete global pre-analysis helps answer the question ``is the jump at the end
of this block really a call?'' based on global behavior, rather than local heuristics,
as in past work.  That is, the global pre-analysis simply simulates the static analysis
up to bounded work and summarizes the observed behaviors. We filter out the spurious \predname{PrivateCallAndContinuation} inferences
by ensuring
the variable holding the value of the continuation is used as a jump target for \emph{some} block and context,
as shown in the rule below (with \predname{Pre\_PrivateCallAndContinuation} being the original definition):

\begin{datalogcode}
  PrivateCallAndContinuation(block, continuation):-
    Pre_PrivateCallAndContinuation(block, continuation),
    BlockPushesContinuation(block, contVar, continuation),
    BlockJumpTarget(_, _, contVar, continuation).
\end{datalogcode}


If our global analysis work limit is too small, causing some valid execution behaviors to not be modeled,
some valid \predname{PrivateCallAndContinuation} facts could be considered spurious
and be filtered out. This could negatively affect the precision and scalability of the
main global analysis: the context abstraction could be missing elements, collapsing together behaviors
that would be best kept distinct. The rest of our pipeline will remain unaffected.
It is up to our experimental evaluation to validate that the incompleteness
of the global analysis does not practically limit the effectiveness of the main analysis.

\subsection{Removing Spurious PublicCall facts}

We use the output of the incomplete global pre-analysis to filter out any spurious
\predname{PublicCall} inferences similarly to how we filtered spurious \predname{PrivateCallAndContinuation}
facts in the previous subsection.

In the case of \predname{PublicCall} facts we introduce a relation that tracks the variables
that hold the function selector bytes, and ensure that one of these variables is used in a comparison
with a small (<= 4 bytes) value.

\subsection{Recovering Important Edges}

The most interesting application of the global pre-analysis is
for identifying imprecision-introducing edges in a \emph{generic, agnostic} way.
This is captured in the definition of predicate \predname{ImportantEdge}, in Figure
\ref{fig:importantEdges}.  Effectively, the logic offers a generic way
to assess that a block is crucial for precision: it is reached with
the stack contents having multiple distinct values (at some position),
and none of its predecessors have this property upon their exit. That
is, the block is reached by multiple predecessors, each establishing
their own run-time conditions. Thus, the edge reaching the
block is important for precision, and therefore worth remembering in the
static context.

\begin{figure}
\begin{datalogcode}
  ImpreciseBlockInput(ctx, block, index):-
    BlockInput(ctx, block, index, var1),
    BlockInput(ctx, block, index, var2),
    var1 != var2.

  ImpreciseBlockOutput(ctx, block, index):-
    BlockOutput(ctx, block, index, var1),
    BlockOutput(ctx, block, index, var2),
    var1 != var2.

  ImpreciseBlockInputFromPrevious(ctx, block, index):-
    ImpreciseBlockInput(ctx, block, index),
    GlobalBlockEdge(prevCtx, prevBlock, ctx, block),
    ImpreciseBlockOutput(prevCtx, prevBlock, index).

  ImprecisionIntroducedAtEdge(fromCtx, fromBlock, toCtx, to, index):-
    GlobalBlockEdge(fromCtx, fromBlock, toCtx, to),
    ImpreciseBlockInput(toCtx, to, index),
    !ImpreciseBlockInputFromPrevious(toCtx, to, index),
    !ImpreciseBlockOutput(fromCtx, fromBlock, index).

  ImportantEdge(from, to):-
    ImprecisionIntroducedAtEdge(_, from, _, to, _).
\end{datalogcode}
\caption{Logic identifying imprecision introducing block edges}
\label{fig:importantEdges}
\end{figure}

The introduction of \predname{ImportantEdge} requires additions to our \funcname{Merge} constructor, which can be
found in Figure~\ref{fig:context-sensitivity-rules-v2}.

\begin{figure}
  \begin{tabular}{l}
  $B$: set of basic blocks \\
  $PC$: set of private contexts, \args{PC} $\cong$ $B^n$ \\
  $C$: set of contexts, \args{C} $\cong$ \args{B $\times$ PC}, as labeled record \record{\args{B}}{\args{PC}}\\
  \\
  Initially, \args{ctx} = \record{\textsc{Null}}{[]}\\
\funcname{Merge}(\record{\args{u}}{\args{p}}, \args{cur}, \args{next}) =
$\begin{cases}
    \record{\args{next}}{p}, \ \ \text{if \pred{PublicCall}{cur, next}}\\
    \\[-1em]

    \record{\args{u}}{[\args{cur}, \predname{First}_{n-1}(\args{p})]},\\
    \begin{tabular}{l} \text{if \pred{PrivateCallAndContinuation}{cur, *}} \\
      \text{or (\pred{PrivateReturn}{cur}} \\
    \ \ \ \  \text{and ($\nexists c$ $\in \args{p}$: \pred{PrivateCallAndContinuation}{c, next})}\\
      \text{or (\pred{ImportantEdge}{cur, next})}
    \end{tabular} \\

    \record{\args{u}}{\funcname{CutTo}(\args{p}, c)},\\
    \begin{tabular}{l}
      \text{if \pred{PrivateReturn}{cur}}\\
      \ \ \ \  \text{and ($\exists c$ $\in \args{p}$: \pred{PrivateCallAndContinuation}{c, next})}
    \end{tabular} \\
    \\[-1em]
    \record{\args{u}}{\args{p}},  \text{otherwise} \\
  \end{cases}
  $

\end{tabular}
\caption[]{Context constructor for shrinking context
  sensitivity following an incomplete global pre-analysis.}
  \label{fig:context-sensitivity-rules-v2}
\end{figure}

In the case of the pre-analysis incompleteness having a negative impact, we will miss some possible \predname{ImportantEdge} inferences,
without affecting our main global analysis or the rest of our pipeline in any way. Again, this effect will be evaluated
experimentally.

\section{Human Study Details}
\label{app:study}

In this Appendix we'll present the layout of the form completed by the human study participants.

\subsection{Invitation}

You've been invited to participate in a study conducted by researchers at the University of Athens and Dedaub on the quality of EVM bytecode decompilers.
The study evaluates three state-of-the-art industrial and academic decompilers.
You will be assigned 3 tasks, each designed to take 10-20 minutes.

For each task you will be given the source code of a smart contract along with the decompilation outputs of the 3 tools, and a method.
You will be tasked with evaluating the decompilation outputs for the given method and report which of the tools are able to accurately recover the method's logic.

To the extend possible, you should attempt to ignore the differences in decompilation style such as the inference of private functions
vs the inlining of all code, naming conventions, etc.

\subsection{Demographic Questions}

\begin{enumerate}
  \item How many years of experience do you have in EVM smart contract security and/or binary reverse engineering?
  \begin{itemize}
    \item 0--2
    \item 2--4
    \item >=5
  \end{itemize}
  \item Which of the options best describes your role?
  \begin{itemize}
    \item Security Researcher/Auditor
    \item Reverse Engineer
    \item Student
    \item Smart Contract Developer
    \item Other
  \end{itemize}
\end{enumerate}

\subsection{Decompilation Task}

\begin{enumerate}
  \item Please share the details of your assigned task (link to gist and requested method).
  \item Decompiler A was able to to accurately express the logic of the given method.
  \begin{itemize}
    \item Strongly Disagree
    \item Disagree
    \item Neutral
    \item Agree
    \item Strongly Agree
  \end{itemize}
  \item Decompiler B was able to to accurately express the logic of the given method.
  \begin{itemize}
    \item Strongly Disagree
    \item Disagree
    \item Neutral
    \item Agree
    \item Strongly Agree
  \end{itemize}
  \item Decompiler C was able to to accurately express the logic of the given method.
  \begin{itemize}
    \item Strongly Disagree
    \item Disagree
    \item Neutral
    \item Agree
    \item Strongly Agree
  \end{itemize}
\end{enumerate}{}

\end{document}